\documentclass[fleqn, usenatbib]{mnras}

\usepackage{newtxtext,newtxmath}

\usepackage[T1]{fontenc}
\usepackage[normalem]{ulem}

\DeclareRobustCommand{\VAN}[3]{#2}
\let\VANthebibliography\thebibliography
\def\thebibliography{\DeclareRobustCommand{\VAN}[3]{##3}\VANthebibliography}

\usepackage{graphicx}	
\usepackage{amsmath}	
\usepackage[euler]{textgreek}
\usepackage{soul}
\usepackage{tablefootnote}

\title[Digging deeper into NGC\,6868]{Digging deeper into NGC\,6868 I: stellar population}
\author[J. P. V. Benedetti et al.]{
João P. V. Benedetti,$^{1}$ \thanks{E-mail: joao.benedetti@ufrgs.br (JPVB)}
Rogério Riffel,$^{1}$ \thanks{E-mail: riffel@ufrgs.br (RR)}
Tiago Ricci,$^{2}$
Marina Trevisan,$^{1}$
Rogemar A. Riffel,$^{3}$
\newauthor
Miriani Pastoriza,$^{1}$
Luis G. Dahmer-Hahn,$^{4}$
Daniel Ruschel-Dutra,$^{5}$
Alberto Rodríguez-Ardila,$^{6,7}$ \newauthor
Jose A. Hernandez-Jimenez$^{8}$ and
João Steiner$^{9}$ \thanks{In Memorian.}
\\
$^{1}$Departamento de Astronomia, Universidade Federal do Rio Grande do Sul. Av. Bento Gonçalves 9500, 91501-970 Porto Alegre, RS, Brazil\\
$^{2}$Universidade Federal da Fronteira Sul, 97900-000 Campus Cerro Largo, RS, Brazil\\
$^{3}$Departamento de Física, Universidade Federal de Santa Maria, Centro de Ciências Naturais e Exatas, 97105-900 Santa Maria, RS, Brazil\\
$^{4}$Shanghai Astronomical Observatory, Chinese Academy of Sciences, 80 Nandan road, Shanghai 200030, China\\
$^{5}$Departamento de Física - CFM - Universidade Federal de Santa Catarina, 476, 88040-900 Florianópolis, SC, Brazil\\
$^{6}$Laboratório Nacional de Astrofísica/MCT - Rua dos Estados Unidos 154, Bairro das Nacões. CEP 37504-364 Itajubá, MG, Brazil \\
$^{7}$INPE - Instituto Nacional de Pesquisas Espaciais, Av. dos Astronautas, CEP 12227-010, São José dos Campos - SP, Brazil\\
$^{8}$Universidade do Vale do Para\'{\i}ba, Av. Shishima Hifumi, 2911, Zip Code 12244-000, S\~ao Jos\'e dos Campos, SP, Brazil \\
$^{9}$Instituto de Astronomia, Geofísica e Ciências Atmosféricas, Universidade de São Paulo, 05508-900 São Paulo, Brazil 
}

\date{Accepted XXX. Received YYY; in original form ZZZ}

\pubyear{2022}

\begin{document}
\label{firstpage}
\pagerange{\pageref{firstpage}--\pageref{lastpage}}
\maketitle

\begin{abstract}

We use Gemini integral field unit observations to map the stellar population properties in the inner region ($\sim680\times470$~{pc\textsuperscript{2}}) of the galaxy NGC\,6868. In order to understand the physical and chemical properties of the stellar content of this galaxy, we performed stellar population synthesis using the {\sc starlight} code with the MILES simple stellar population models. We measured the absorption line indices Fe4383, Mg\textsubscript{2}, Mg\textsubscript{b}, Fe5270, Fe5335 for the whole FoV, and used them to derive Fe3 and [MgFe]'. These indices were used to derive [\textalpha/Fe]. This galaxy is dominated by old metal-rich populations (12.6 Gyr; 1.0 and 1.6 Z$_\odot$) with a negative metallicity gradient. We also found a recent ($\sim63$~{Myr}) metal-rich (1.6~{Z$_{\odot}$}) residual star formation in the centre of the galaxy. A dust lane with a peak extinction in the V band of 0.65~{mag} is seen. No signs of ordered stellar motion are found and the stellar kinematics is dispersion dominated. All indices show a spatial profile varying significantly along the FoV. Mg\textsubscript{2} shows a shallow gradient, compatible with the occurrence of mergers in the past. Mg\textsubscript{b} and Fe3 profiles suggest different enrichment processes for these elements. We observe three distinct regions: for R<100~pc and R>220~pc, Mg\textsubscript{2}, Mg\textsubscript{b} anti correlate with respect to Fe3 and [MgFe]', and for 100~pc<R<220~pc, they correlate, hinting at different enrichment histories. The [\textalpha/Fe] profile is really complex and has a central value of $\sim 0.2$~{dex}. We interpret this as the result of a past merger with another galaxy with a different [\textalpha/Fe] history, thus explaining the [\textalpha/Fe] maps.

\end{abstract}

\begin{keywords}
galaxies: individual (NGC\,6868), galaxies: nuclei, galaxies: elliptical and lenticular, cD, galaxies: stellar content
\end{keywords}



\section{Introduction}
\label{sec:int}
    
    Galaxies can widely be defined as either {\it passive}, that are not actively forming stars and host a red and old stellar population, and {\it star-forming} galaxies, that are blue, hosting large fractions of young stellar populations. This bi-modality was found in many studies over the years \citep[e.g.][]{KauffmannEtAl2003, BaldryEtAl2004, NoeskeEtAl2007, WetzelEtAl2012, vanderWelEtAl2014}, even at high redshifts \citep[$z > 2.5$][for example]{BrammerEtAl2009, MuzzinEtAl2013}. However, it is not yet clear which mechanisms are regulating star formation and transforming the blue galaxies into {\it red-and-dead} ones. A major challenge in modern astrophysics is to determine the physical mechanism acting in quenching star formation in galaxies.
    
    Nowadays, it is established that active galactic nuclei (AGN) feedback plays an important role in regulating the star formation (SF) of its host galaxy \citep[][]{DiMatteoEtAl2005, HopkinsElvis2010, Harrison2017, Storchi-BergmannSchnorr-Muller2019, RiffelEtAl2021, EllisonEtAl2021}. The gas inflowing, responsible for SF, also feeds the supermassive black hole (SMBH), triggering the AGN episode that can either heat (or expel) the gas, thus shutting down the SF \citep[][]{Fabian2012, KingPounds2015, ZubovasBourne2017, TrusslerEtAl2020}. In cosmological simulations without including AGN and supernova (SN) feedback, the observed luminosity function of galaxies cannot be reproduced: the larger and smaller galaxies end up with higher masses than observed in the present-day universe \citep[][]{SpringelEtAl2005}. Also, the ages of the stars from the most massive galaxies are underestimated when compared with observations \citep[][]{CrotonEtAl2006}. These results show that some form of gas striping or heating must be taking place in these objects. However, distinguishing the nature of such processes is still challenging, once the simulations cannot reach the physical scales involved and use \textit{ad hoc} prescriptions to account for these mechanisms \citep[][]{SchayeEtAl2015}. In order to really disentangle the effects of SN feedback and AGN we need to look at the vicinity of SMBH and trace the star formation history (SFH) of that population in order to understand the effect of the AGN in the stellar population \citep[][]{RiffelEtAl2021}.
    
    Past studies have tried to establish this link, but the results have been controversial. Despite SF being common in AGNs \citep[][]{RiffelEtAl2009,Ruschel-DutraEtAl2017, MallmannEtAl2018, RiffelEtAl2021, BurtscherEtAl2021,Dahmer-HahnEtAl2021, RiffelEtAl2022}, the time scale for starting the star formation \citep[$\sim100$~{Myr},][]{HickoxEtAl2014, BurtscherEtAl2021} is far greater than for the AGN triggering \citep[$\sim0.1-1$~{Myr},][]{NovakEtAl2011, SchawinskiEtAl2015}, preventing us from connecting the two. Some studies show a correlation between the fraction of young populations and the AGN luminosity, where the most luminous sources present the highest fraction \citep[][]{RiffelEtAl2009, Ruschel-DutraEtAl2017, ZubovasBourne2017, MallmannEtAl2018}. However, the hard X-ray (14-195~{keV}) luminosity from the galaxies does not seem to correlate with the fraction of young populations \citep[][]{BurtscherEtAl2021}. Instead, mass loss from intermediate-age stars seems to be important in AGN feeding \citep{RiffelEtAl2022}.
    
    
    Most of the previous studies have been done on relatively bright objects. However, the most common form of AGN in the local Universe is low luminosity AGN (LLAGN) in massive galaxies \citep[][]{Ho2008}. Most of them are classified as low-ionization nuclear emission-line region (LINER) objects \citep[][]{Heckman1980}. However, despite their significance, the physical nature of these objects is still poorly understood. Since their discovery, many mechanisms have been proposed to explain the LINER signature beyond the LLAGN paradigm \citep[][]{FerlandNetzer1983, HalpernSteiner1983}, once many other physical processes can mimic the same spectral signatures without the LLAGN \citep[these objects are known as LIERs][]{CidFernandesEtAl2011, BelfioreEtAl2016} such as shocks \citep[][]{Heckman1980}, hot low-mass evolved stars (HOLMES) \citep[][]{BinetteEtAl1994, YanBlanton2012, PapaderosEtAl2013, SinghEtAl2013, BelfioreEtAl2016, OliveiraEtAl2022} and starbursts with ages between 3 and 5 Myr, dominated by Wolf-Rayet stars \citep[][]{BarthShields2000}. With current observational technology, one can disentangle the different ionization mechanisms by performing detailed spatially resolved studies analysing both the stellar population and the ionized gas components.
    
    Even in the LLAGN hypothesis, the effects of such objects in their host galaxy are still uncertain as most studies focusing on the AGN impact over galaxies are performed with high-luminosity AGN \citep[e.g. Seyferts and quasars, e.g.][]{NayakshinZubovas2012}. With the rising importance of LINERs, new studies have been analysing such impacts, although a complete picture is yet to be drawn and further research is needed.
    
    
    Integral Field Unit (IFU) spectroscopy has expanded our view towards early-type galaxies, their formation, and evolution. This technique allows one to perform spatially resolved studies of stellar populations and better constrain the kinematical structure of these objects, with, for example, the emergence of kinematically distinct cores (KDCs), counter-rotating stellar discs \citep[see][for a review]{Cappellari2016}. Despite previous studies being able to reproduce the stellar population parameters of ETGs with a rapid in-situ conversion of gas into stars \citep[e.g.][]{ChiosiCarraro2002} including a fully consistent chemical evolution \citep[e.g.][]{VazdekisEtAl1997}, the emergence of these structures has been seen as evidence for the importance of merger processes in ETG formation. Dry minor mergers have already been established as a subsequent growth pathway for ETGs \citep[the two phases of galaxy formation][]{OserEtAl2010, Navarro-GonzalezEtAl2013}, however, they rarely affect the central regions of galaxies. Therefore, studying these structures in galactic cores may help us further elucidate the formation and evolution of ETGs \citep[e.g.][]{KrajnovicEtAl2015}.
        
\begin{figure*}
    \centering
    \includegraphics[width=0.95\textwidth,height=0.95\textheight, keepaspectratio]{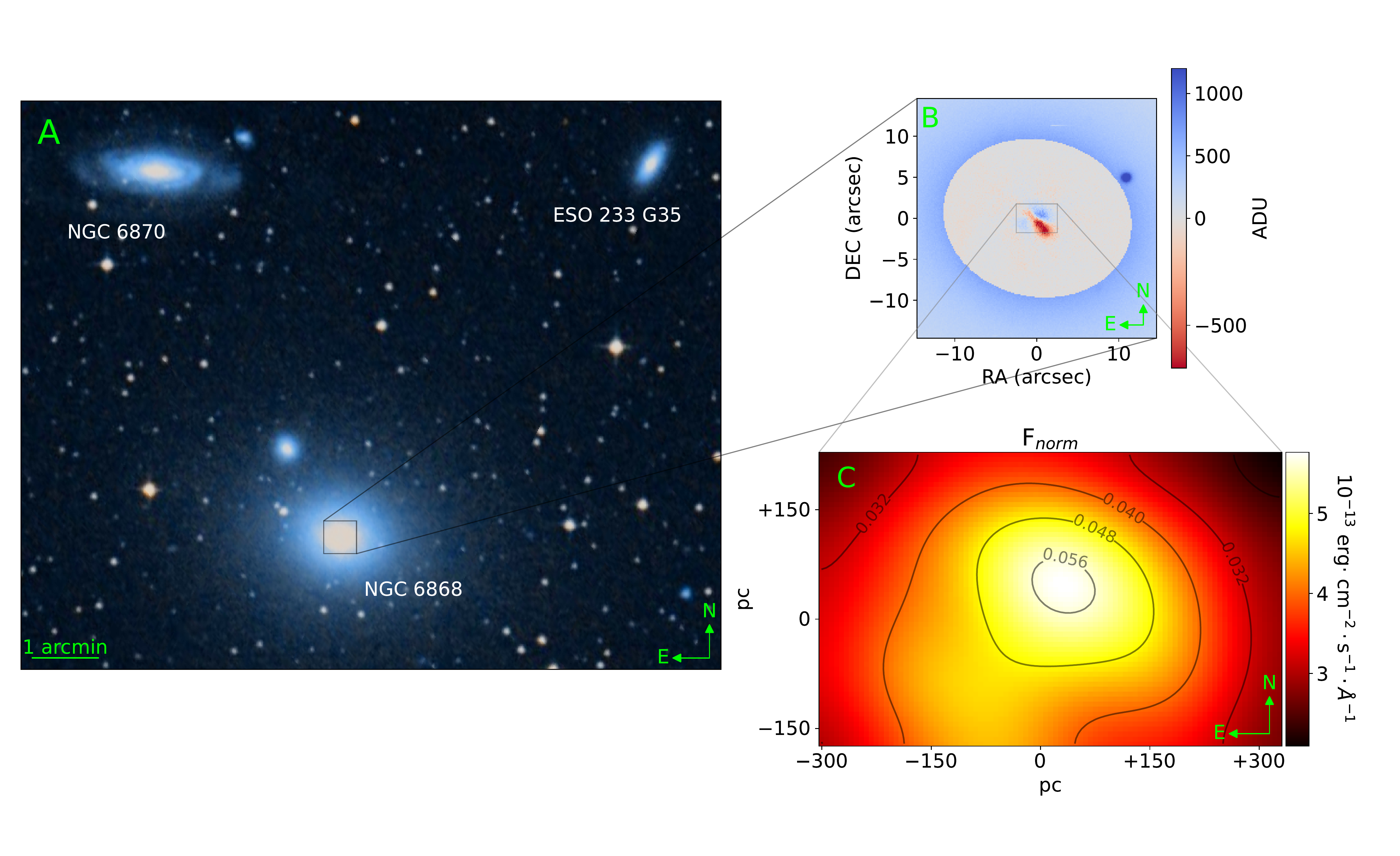}
    \caption{Images from NGC\,6868 in three different scales. (a) Composite DSS image showing NGC\,6868 and close neighbours. It is the brightest group member from the Telescopium group (AS0851). (b) Residuals from the photometric modelling of the acquisition image in the r band. A clear dust lane, also reported by \citet[][]{Veron-CettyVeron1988}, is seen in the centre of NGC\,6868. At $\sim$11"W and 5"N a stellar-like object is found and, according to \citet[][]{HansenEtAl1991}, may be a cannibalised galaxy from which spiral features are emerging. (c) Continuum image from NGC\,6868 extracted from our GMOS data cube at 5700~{\AA} exhibiting an irregular light profile with a distortion in the SE direction. The (0,0) is defined as the peak in the continuum image corrected by extinction, as can be seen in Fig.~\ref{fig:fnorm_corr}.}
    \label{fig:big_picture}
    \end{figure*}

    From the above, detailed studies thoroughly analysing the nuclear regions of galaxies probing the stellar population and gas content are fundamental to elucidate the impact of the AGN with respect to the galaxy evolution. In this sense, an "artisanal" approach is better at analysing the details that would otherwise be missed in large surveys. With this in mind, here we present a detailed GMOS IFU study of the galaxy NGC\,6868, a nearby \citep[$27.70$~{Mpc},][]{TullyEtAl2013} elliptical galaxy \citep[E2,][]{deVaucouleursEtAl1991}. Some basic parameters extracted from NED can be seen in table \ref{tab:basic_pars} and three images from NGC\,6868 in different scales are shown in Fig.~{\ref{fig:big_picture}}. NGC\,6868 is the brightest member of the Telescopium group. 
    \citet{RickesEtAl2008} have shown that NGC\,6868 exhibits LINER emission in its centre, which has been attributed to a combination of photoionization by an LLAGN and shocks. They also investigate this galaxy's metallicity distribution and ionized gas by means of long-slit spectroscopy and stellar population synthesis. According to the authors, Lick indices present a negative gradient indicating an overabundance of Fe, Mg, Na and TiO in the central parts with respect to the external regions. Mg\textsubscript{2} correlates with Fe5270 and Fe5335, suggesting that these elements probably underwent the same enrichment process in NGC\,6868. The lack of correlation between computed galaxy mass and the Mg\textsubscript{2} gradient suggests that this elliptical galaxy was formed by merger events. The stellar population synthesis shows the presence of at least two populations with ages of 13 and 5~{Gyr} old. The fact that this galaxy apparently has multiple ionization scenarios and also shows signs of complex star formation history makes NGC\,6868 an excellent candidate to further investigate the mechanisms behind LINER emission and the processes involved in the evolution of early-type galaxies.
    
    \begin{table}
    \centering
    \caption{Table showing some basic parameters of the galaxy NGC\,6868.}
    \label{tab:basic_pars}
    \begin{tabular}{lc}
    \hline
    Parameter                                                                                               & NGC 6868                                                                                                                                     \\ \hline
    RA (J2000)                                                                                              & 20\textsuperscript{h}09\textsuperscript{m}54\textsuperscript{s}.07                                                                           \\
    Dec. (J200)                                                                                             & -48$^{\circ}$22´46.4´´                                                                                                                        \\
    Morphology\textsuperscript{a}                                                                                              & E2                                                                                                                                           \\
    R (mag)\textsuperscript{b}                                                                                                & 7.91                                                                                                                                         \\
    M$_\text{R}$ (mag)\textsuperscript{b}                                                                                             & -24.7                                                                                                                                        \\
    Diameter (kpc)\textsuperscript{c}                                                                                            & 73.0                                                                                                                                        \\
    L$_\text{X}$ (erg\,s$^{-1}$)\textsuperscript{d}                                                                             & $8.54\cdot10^{40}$                                                                                                                                    \\ 
    Nuclear Activity\textsuperscript{e}                                                                                           & LINER                                                                                                                                    \\ 
    Radio classification\textsuperscript{f}                                                                                    &  Flat-Spectrum Radio Source	                                                                                                                                \\ 
    A$_\text{V}$ (mag)\textsuperscript{g}                                                                                      & 0.152                                                                                                                                        \\
    Radial Velocity (km\,s$^{-1}$)\textsuperscript{h}                                                                            & 2854                                                                                                                                         \\
    Distance (Mpc)\textsuperscript{i}                                                                                          & 27.70                                                                                                                                        \\
    Redshift (z)\textsuperscript{h}                                                                                            & 0.00952                                                                                                                                      \\
    Velocity dispersion (km$\cdot$s$^{-1}$) \textsuperscript{j}                                                                                            & 250                                                                                                                                      \\\hline
    \multicolumn{2}{l}{Data available in NED\tablefootnote{The NASA/IPAC Extragalactic Database (NED) is operated by the Jet Propulsion Laboratory, California Institute of Technology, under contract with the National Aeronautics and Space Administration}}\\
    \textsuperscript{a}\citet{deVaucouleursEtAl1991} & \textsuperscript{b}\citet{CarrascoEtAl2006}\hfill\vadjust{}\\
    \textsuperscript{c}\citet{LaubertsValentijn1989} & \textsuperscript{d}\citet{BabykEtAl2018}\hfill\vadjust{}\\
    \textsuperscript{e}\citet{RickesEtAl2008} & \textsuperscript{f}\citet{HealeyEtAl2007}\hfill\vadjust{}\\
    \textsuperscript{g}\citet{SchlaflyFinkbeiner2011} & \textsuperscript{h}\citet{RamellaEtAl1996}\hfill\vadjust{}\\
    \textsuperscript{i}\citet{TullyEtAl2013} & \textsuperscript{j}\citet{WegnerEtAl2003}\hfill\vadjust{}\\

    \end{tabular}
    \end{table}
    
    NGC\,6868 was already observed in different wavelengths. \citet{MachacekEtAl2010} using X-ray data found strong evidence of a past encounter between NGC\,6868 and NGC \,6861, displaying tidal tails and shells. Moreover, they found X-ray cavities, indicative of past AGN activity. \citet{HansenEtAl1991} studied NGC\,6868 using CCD images and an International Ultraviolet Explorer (IUE) low-resolution spectrum, and they found a dust lane in the centre of the galaxy with an extended dust component with spiral features. A series of papers have reported the presence of ionized gas, finding a disturbed morphology and complex kinematics for the galaxy with a possible counter-rotating disc \citep{BusonEtAl1993, ZeilingerEtAl1996, MacchettoEtAl1996}. \citet{CaonEtAl2000} have analysed long-slit observations with distinct position angles (PAs). They reported a rotating disc of ionized gas, in agreement with the stellar velocity field. However, in a different PA, a counter-rotating gas disc with respect to the stars is found, displaying an inner component that shows a counter-rotation. Also, a KDC was seen where the kinematical break radius is at 3''. \citet{BregmanEtAl1998} using IRAS data confirmed the presence of cold dust. NGC\,6868 has been observed in the radio by \citet{SleeEtAl1994} at  2.3, 5, and 8.4~{GHz} and \citet{MauchEtAl2003} at 843 MHz and \citet{HealeyEtAl2007} reported a low-power flat spectrum radio source in its centre ($\alpha \sim 0.07$). The brightness, temperature and spectral slope are inconsistent with HII regions, so an AGN was the most likely source of the radio emission. \citet{RoseEtAl2019} using ALMA observations detected molecular gas in the centre of NGC\,6868 drifting in non-circular motions. Also, they reported an HI absorption.
    
    In this paper, the first of a series aimed at studying in detail this object, we will focus on the stellar content of NGC\,6868. It is organized as follows: in \S~\ref{sec:data}, we describe the observations and the reduction procedures; in \S~\ref{sec:methods}, we present the employed methodology; in \S~\ref{sec:results}, the results are presented and a comparison with data from other studies. Discussion of the results is made in \S~\ref{sec:disscuss} and the conclusion and summary are made in \S~\ref{sec:conclusion}. Throughout this paper, we assume that solar metallicity corresponds to $\text{Z}_\odot=0.019$ \citep[][]{GirardiEtAl2000}.

\section{Observation and Data Reduction} \label{sec:data}

    NGC\,6868 was observed on 2013 May 04 with the Gemini South Telescope using the Gemini Multi-Object Spectrograph (GMOS) in the IFU mode \citep{Allington-SmithEtAl2002, HookEtAl2004}. This object is part of the DIVING\textsuperscript{3D} survey, which made observations of the central regions of all 170 galaxies in the Southern hemisphere with $B < 12.0$ and $|b| > 15^{\circ}$ (see \citealt{SteinerEtAl2022} for more details). The one slit set-up was used for the observations, resulting in an FoV of 5.0 × 3.5 arcsec\textsuperscript{2}. The B600-G5323 grating was used with a central wavelength of 5620~{\AA} and a spectral range from 4260~{\AA} to 6795~{\AA}. The spectral resolution is 1.8~{\AA}, estimated with the O {\sc i} $\lambda$5577{\AA} sky line. Flat-field exposures, bias and CuAr lamp spectra were acquired for calibration and correction purposes. The seeing of observation was estimated using stars that are present in the acquisition image of the galaxy, taken with the GMOS imager in the r-band (SDSS system). Moreover, the DA white dwarf EG 274 \citep{HamuyEtAl1992} was observed in order to perform the spectrophotometric calibration. These and some other basic information regarding the observation are given in table \ref{tab:pars_obs}.
    
    Standard {\sc IRAF} procedures \citep{Tody1986, Tody1993} were followed to reduce the data using the tasks contained in the {\sc Gemini IRAF} package. Bias, flat-fields, wavelength calibration, dispersion correction, and flux calibration procedures were applied to the science data. To remove cosmic rays, we used the {\sc lacos} software \citep{vanDokkum2001}. The data cube was constructed with a spatial sampling of 0.05 arcsec.
    
    After the standard reduction procedures, other data treatments were applied by means of improving the visualization of the data as described in \citet{MenezesEtAl2019}: removal of the high spatial noise using a Butterworth filter, the correction of the differential atmospheric refraction (DAR), instrumental fingerprint removal through PCA Tomography and Richardson-Lucy deconvolution. 
    
    The removal of high-frequency noise from the spatial dimension was performed by convolving each image of the data cube with a Butterworth filter \citep{GonzalezWoods2008, RicciEtAl2014a}. The filter order used was $n=2$ and the cut-off frequency was $\text{F}_\text{c}=0.14\ \text{F}_{\text{Ny}}$ where $\text{F}_{\text{Ny}}$ is the Nyquist frequency, corresponding to 0.5 spaxel$^{-1}$. This cut-off frequency was chosen to remove only spatial frequencies higher than the PSF of the data cube, assuring that no valid scientific information was lost in this process.
    
    The correction of the differential atmospheric refraction (DAR) consists of spatially shifting the wavelength planes of the data cube so that the spectrum of a given point in the galaxy occupies always the same position at all wavelengths. The correction of the differential refraction effect on the data cube of NGC\,6868 was performed using the equations from \citet{BonschPotulski1998} and \citet{Filippenko1982}, which assume a plane parallel atmosphere and calculate the displacement of the centroid of the galaxy for each wavelength as a function of the zenith distance, the refraction index and other atmospheric parameters.
    
    The PCA Tomography technique \citep[][and references therein]{SteinerEtAl2009} applies Principal Component Analysis (PCA) to data cubes. This procedure searches for spectro-spatial correlation across a given data cube. The results are the eigenvectors (or eigenspectra), which show the correlations between the wavelengths caused by some physical phenomenon or an instrumental fingerprint, and the tomograms, which correspond to the projection of the data cube onto each eigenvector. This is an orthogonal transform meaning it can be reversed. The eigenvectors are ordered by how much of the variance in the data cube they are able to explain, meaning the first eigenvector explains most of the variance and so on. Using this technique, the instrumental fingerprints may appear as one of the eigenvectors that would otherwise be entangled with the data. This instrumental signature may be isolated by building a data cube containing only this issue. In the end, we subtract this fingerprint from the science data cube.
    
    After the removal of the instrumental fingerprints, the reddening caused by the dust within the Milky Way was corrected using the CCM law \citep{CardelliEtAl1989} and $A_V=0.152$~{mag} \citep[][]{SchlaflyFinkbeiner2011}. The telluric lines were also removed and the spectra were brought to the rest frame using the redshift shown in Table \ref{tab:basic_pars}.
    
    Lastly, the Richardson-Lucy deconvolution \citep{Richardson1972, Lucy1974} is an iterative process that aims at reconstructing the image of the galaxy before its convolution with the PSF when it passes through the atmosphere and the optical apparatus of the telescope. After 10 iterations, the final estimated PSF was 0.71 arcsec, estimated from a spatial profile obtained along the red wing of the broad H\textalpha\, emission. An image from the continuum of NGC 6868, extracted from the final data cube, is shown in Fig.~{\ref{fig:big_picture}}.

    \begin{table}
    \label{tab:pars_obs}
    \centering
    \caption{Table displaying some basic observational parameters}
    \begin{tabular}{lc}
    \hline
    Parameter              & NGC\,6868        \\ \hline
    Observation date       & 2013 May 04      \\
    Gemini Programme       & GS-2013A-Q-52   \\
    Seeing (arcsec)        & 0.77            \\
    Airmass                & 1.056           \\
    T$_{\text{exp}}$~{(s)} & 1800            \\
    Number of exposures    & 1              \\\hline
    \end{tabular}
    \end{table}

\section{Methodology} \label{sec:methods}

    \subsection{Stellar population synthesis} \label{sec:meth_syn}
    
    In order to derive the SFH (star formation history), we used the {\sc starlight} code \citep{CidFernandesEtAl2004, CidFernandesEtAl2005, CidFernandesEtAl2013, CidFernandes2018} which fits the continuum spectra by combining in different proportions the contribution from different simple stellar populations (SSPs), taking into account reddening and kinematical parameters. In other words, it tries to match the observed spectrum ($O_\lambda$) with a modelled one ($M_\lambda$), described by
    \begin{equation}
        M_\lambda = M_{\lambda_0}\left[\sum_{j=1}^{N_*} x_j b_{j,\lambda}r_\lambda\right] \otimes G(v_*,\sigma_*),
        \label{eq:modelspec}
    \end{equation}
    where $M_{\lambda_0}$ is the flux in a predetermined normalization wavelength, $N_*$ is the number of elements in the SSP base, $x_j$ is the j-th component of the population vector ($\vv{x}$) that stores the light contribution from each SSP (with respect to the normalization wavelength\footnote{We normalized our spectra in the 5700~{\AA} region, due to the lack of significant stellar absorption bands and having good S/N in the whole FOV.}, $\lambda_0$ ). $b_{j,\lambda}$ is the spectrum of the j-th component, $r_\lambda$ is the reddening factor, defined by $r_\lambda = 10^{-0.4(A_\lambda - A_{\lambda_0})}$) and $A_\lambda=A_\text{V}q_\lambda$, where $q_\lambda$ is the extinction law evaluated at $\lambda$. Lastly, there is a convolution with a Gaussian distribution to take into account the line-of-sight velocity distribution (LOSVD) of the stellar component in the spectra, where $v_*$ is the line-of-sight stellar velocity and $\sigma_*$ is the stellar velocity dispersion. To determine the best fit, the code tries to minimize a $\chi^2$ defined by 
    \begin{equation}
        \chi^2=\sum_\lambda[(O_\lambda-M_\lambda)\omega_\lambda]^2
        \label{chi2eq}
    \end{equation}
    where $\omega_\lambda$ is the weight. Using this parameter, we are able to mask ($\omega_\lambda=0$) spurious features or contributions from other non-stellar components (e.g. emission lines from the ionized gas) or give more weight to important regions of our spectra (e.g. characteristic adsorptions that allow better kinematical predictions, if that is the intended objective). Along with the emission lines present in our spectra, the \ion{Mg}{i} absorption (an \textalpha\ element) was masked during the synthesis in order to minimize possible degeneracies that can be introduced by the \textalpha-enhancement effects in the determination of the galaxy metallicity (the [\textalpha/Fe] is derived in \S~\ref{sec:meth_idx}).
    
    One of the fundamental ingredients in this method is the SSPs used in the fit. We constructed our base with the models developed by \citet[][hereafter E-MILES]{VazdekisEtAl2016}, using the evolutionary tracks of \citet{GirardiEtAl2000} and the \citet{Kroupa2001} initial mass function. These models were chosen because their wavelength range overlaps with our data and have a better spectral resolution (2.51~\AA~FWHM) when compared to other stellar population models \citep[e.g.][]{BruzualCharlot2003, MarastonStromback2011, ConroyEtAl2009}\footnote{The high resolution is fundamental to precisely modelling the stellar absorptions, allowing a detailed study of the gas kinematics when discounting the stellar component. This allows a self-consistent analysis that will be pursued in a future paper (Benedetti et al., in preparation).}. Moreover, MILES is a modern empirical stellar library spanning a wide range of stellar parameters, therefore, allowing us to better explore different stellar population properties in our object. The final SSPs span 15 ages (0.0631, 0.1, 0.16, 0.28, 0.50, 0.89, 1.26, 1.41, 2.51, 3.98, 6.31, 7.94, 10.0, 11.2, 12.6~Gyr) and six metallicities (0.005, 0.02, 0.2, 0.4, 1.0, 1.6~Z$_\odot$).
    
    The E-MILES models, however, lack really young and hot stars, having only SSPs with ages greater than $63$~{Myr}. In order to assess the possibility of such a young population in NGC\,6868, we performed stellar synthesis with the \citet{BruzualCharlot2003} models, which include stars as young as 0.1 Myr in the whole FoV. We found no contribution from components with less than $63$~{Myr}.
    
    Since this object has no ongoing star formation, we decided to use the reddening law from \citet{CardelliEtAl1989} to model the dust attenuation (A$_V$) intrinsic to our object. To account for a possible featureless continuum (FC) emission of an AGN, we followed \citet{RiffelEtAl2009} and a power-law spectrum with $f \sim \nu^{-1.5}$ was added to the base. In other test runs, we have included FC with different exponents, ranging from -1.75 to -1.0, however, no significant contribution from any of these components was found.
    
    Finally, to better understand the age and metallicity spatial distribution, we calculated the light-weighted mean stellar age ($\langle t \rangle_L$), as
    \begin{equation}
        \langle t \rangle_L=\sum_j x_j\log(t_j),
        \label{eq:mean_age}
    \end{equation}
    and the light-weighted mean stellar metallicity\citep[$\langle Z \rangle_L$, ][]{CidFernandesEtAl2005}, as 
    \begin{equation}
        \langle Z \rangle_L=\sum_j x_jZ_j.
        \label{eq:mean_z}
    \end{equation}
    $\log(t_j)$ is used in the computation of $\langle t \rangle_L$ because the ages of the stellar population span many other orders of magnitude from $10^6$ to $10^{10}$.

    To optimize the data management, we used the {\sc megacube} tool \citep{MallmannEtAl2018, RiffelEtAl2021} which takes as input the data cube and prepares the data for the synthesis procedure, executes the synthesis and also performs the preliminary analysis (e.g. calculate the equations \ref{eq:mean_age} and \ref{eq:mean_z}) as well as mounting the maps with the important parameters.

    \subsection{Indices measurements and Alpha-enhancement} \label{sec:meth_idx}
    
    To better constrain the assembly history of NGC\,6868, especially the \textalpha-enhancement, we have measured indices for the absorption lines. We measured the indices for Fe4383, Mg\textsubscript{2}, Mg\textsubscript{b}, Fe5270, Fe5335 using the definitions presented in \citet{RiffelEtAl2019} which are based on \citet{WortheyEtAl1994} and were subsequently used to derive Fe3\footnote{$\text{Fe}3 = (\text{Fe}4383 + \text{Fe}5270 + \text{Fe}5335)/3$ \citep{Kuntschner2000}} and [MgFe]'\footnote{$[\text{MgFe}]'=\sqrt{\text{Mg}_\text{b}\left(0.72\times\text{Fe}5270 + 0.28\times\text{Fe}5335\right)}$ \citep[][]{ThomasEtAl2003}}. All spaxels were corrected to the rest frame using the line-of-sight Doppler shift velocity derived by {\sc starlight}. At first, we decide not to correct the effects due to the velocity dispersion.
    
    The in-house  {\sc pacce} code \citep{RiffelBorgesVale2011} was used to perform the equivalent widths (EW) measurements of these indices. The code uses predefined continuum bands around a given line and fits a pseudo-continuum line. Once the line is fitted, the EW is calculated through
    
    \begin{equation}
        W_\lambda=\left(1-\frac{A_2}{C}\right)\left(\lambda_u-\lambda_i\right)
    \end{equation}
    being $W_\lambda$ the measured EW of the line, $A_2$ and $C$ are the integrated areas below the absorption and below the pseudo-continuum, respectively, and $\lambda_u$ and $\lambda_i$ are the predetermined final and initial wavelength of the absorption feature. The trapezium integration method is used and the calculation of W$_\lambda$ is iterated over the whole cube. 
    
    A problem we faced during these measurements was the contamination from the weak [\ion{N}{i}] $\lambda~5199$~{\AA} emission line which is located between the \ion{Mg}{i} absorption and the pseudo continuum definition of Mg\textsubscript{b}. We were able to work this around by modelling the emission line profile and subtracting this contribution from the spaxels. In order to properly model this component, we discounted the previously derived stellar spectra and fitted the remaining continuum with a high-order polynomial. Afterwards, we adjusted this line using the {\sc ifscube} package \citep{Ruschel-DutraOliveira2020}. We modelled the [\ion{N}{i}] $\lambda~5199$~{\AA} emission line using two Gaussian components and coupled its kinematics with the [\ion{N}{ii}] $\lambda\lambda 6548, 6583~$~{\AA}. Additional details on emission line fitting will be presented in Benedetti et al. ({\it in preparation}). Once the lines were properly fitted we were able to discount this emission and remeasure the \ion{Mg}{i} indices.
    
\begin{figure}
    \centering
    \includegraphics[width=0.95\columnwidth, height=0.95\textheight, keepaspectratio]{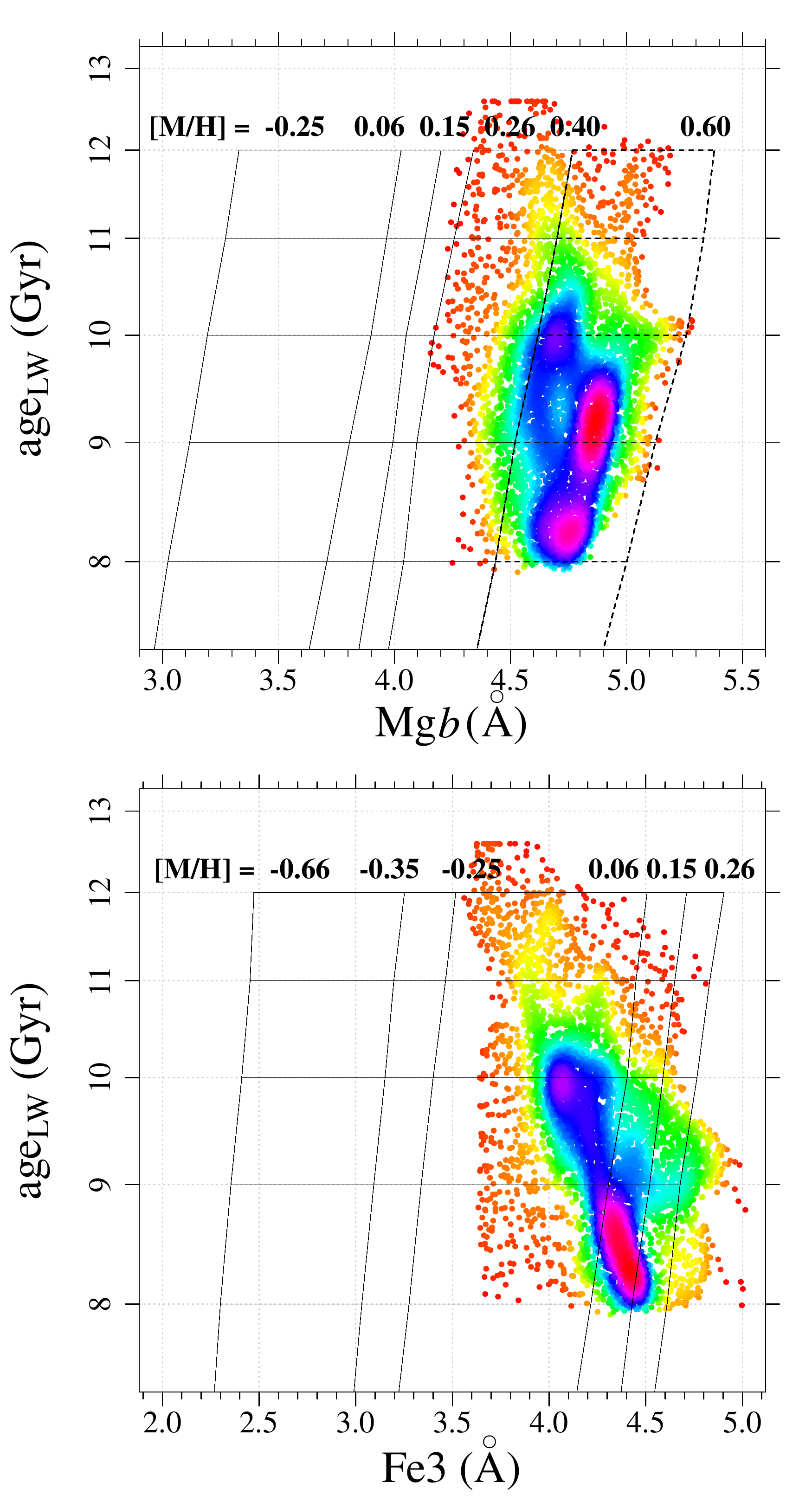}
    \caption{Diagrams showing the grids used to derive the Z\textsubscript{Mg$_b$} (top) and the Z\textsubscript{Fe3} (bottom) using the light-weighted mean age derived from {\sc starlight} and the measured EW from each absorption.Each spaxel is represented by one dot.}
    \label{fig:grids}
    \end{figure}

    Aimed at constraining the assembly history of NGC\,6868, we derived the [\textalpha/Fe] of the stellar population. Following the approach described in \citet{LaBarberaEtAl2013} using the indices Mg$_\text{b}$, Fe3 and the luminosity-weighted age derived from {\sc starlight} to obtain Z\textsubscript{Mg$_\text{b}$} and Z\textsubscript{Fe3}. To measure the Mg$_\text{b}$ and Fe3 indices, we first broadened our spectra convolving with a Gaussian to match the spectral resolution of {\sc MILES} models \citep[2.51~\AA~FWHM,][]{VazdekisEtAl2015} and the measured indices were corrected by the velocity dispersion following the prescriptions by \citet[][]{delaRosaEtAl2007}. We then interpolated the \citet[][]{VazdekisEtAl2015} models grids by fixing the stellar population age to obtain Z\textsubscript{Mg$_\text{b}$} and Z\textsubscript{Fe3}, as illustrated in Fig.~{\ref{fig:grids}}. As mentioned in \citet{LaBarberaEtAl2013}, one may need to extrapolate the grids in \textalpha-enhanced populations. This is represented as the dotted lines in the same figure. Afterwards, we calculated the proxy [Z\textsubscript{Mg$_b$}/Z\textsubscript{Fe3}]= Z\textsubscript{Mg$_b$}-Z\textsubscript{Fe3} which finally can be used to get \citep[][]{VazdekisEtAl2015}:
    \begin{equation}
        [\alpha/\text{Fe}]=0.02+0.56[Z\textsubscript{Mg$_b$}/Z\textsubscript{Fe3}].
        \label{eq:aFe}
    \end{equation}
    
\section{Results} \label{sec:results}

    \subsection{Stellar population synthesis} \label{sec:res_synth}
    
\begin{figure*}
    \centering
    \includegraphics[width=0.8\textwidth, height=0.95\textheight, keepaspectratio]{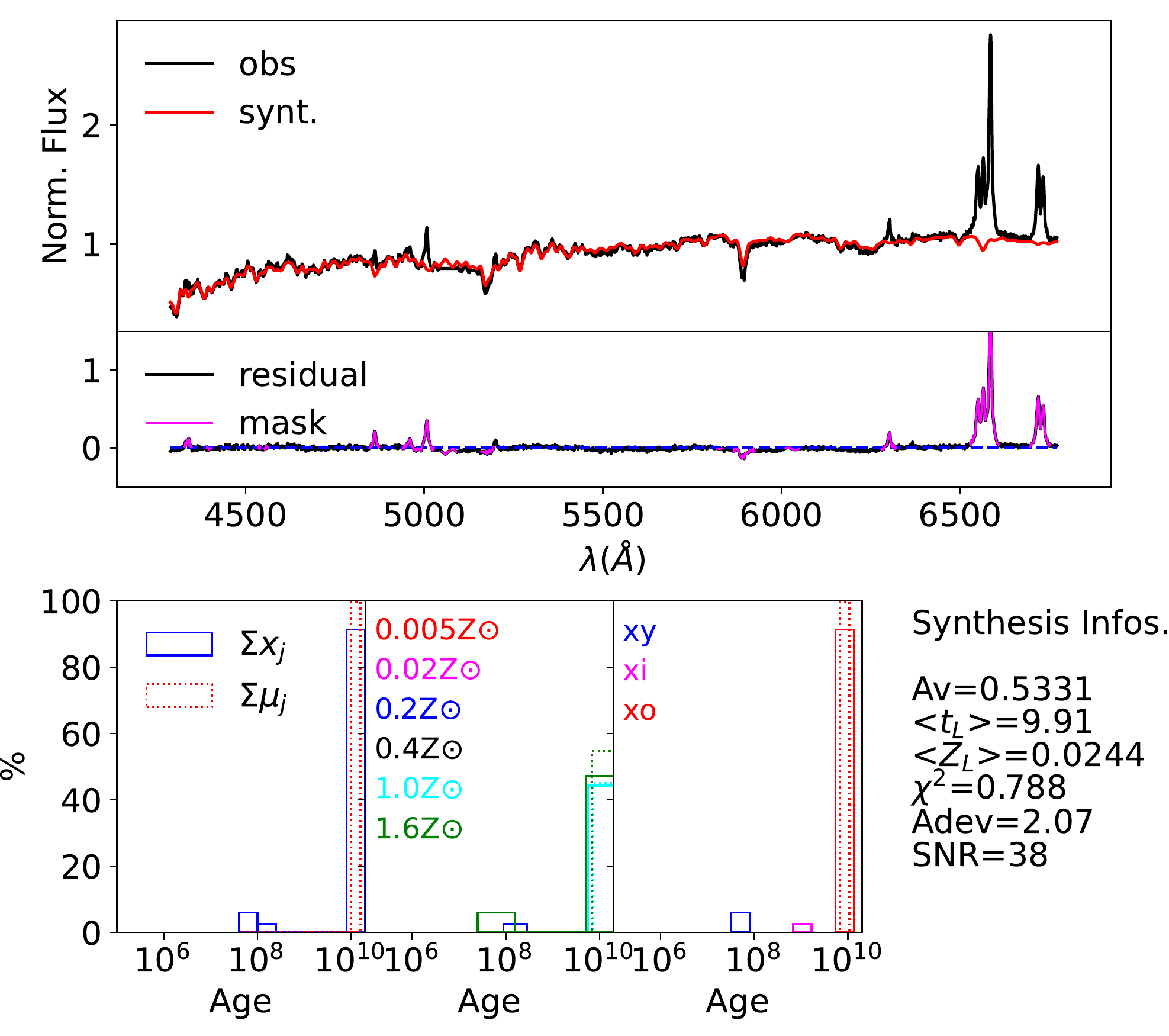}
    \caption{\textbf{Top:} Example spectra extracted from the spaxel with the highest continuum flux. The observed spectra (black) and the synthesised spectrum (red) are shown as well as the residuals (blue) and the masked pixels (pink). \textbf{Bottom:} Histograms showing the stellar population composition found by {\sc starlight}. Continuous lines show light-weighted parameters and dashed lines, mass-weighted parameters. From left to right, the sum of contributions of components with the same age but different metallicities; contribution from each SSP in the base; age and metallicities contribution summed in predefined age bins, being  {\sc xi} SSPs between 0.1~{Myr} until 100~{Myr}, {\sc x}, 100~{Myr} until 2~{Gyr} and {\sc xo}, 2~{Gyr} until 13.7~{Gyr}. Further right are other parameters derived by the synthesis: The reddening in the V band, light- and mass-weighted mean age, light- and mass-weighted mean metallicity, $\chi^2$, Adev and the S/N measured between 5670-5730~\AA. The code clearly indicates the presence of old populations followed by a small fraction of young populations.}
    \label{fig:example}
    \end{figure*}

\begin{figure*}
    \centering
    \includegraphics[width=0.9\textwidth, height=0.95\textheight, keepaspectratio]{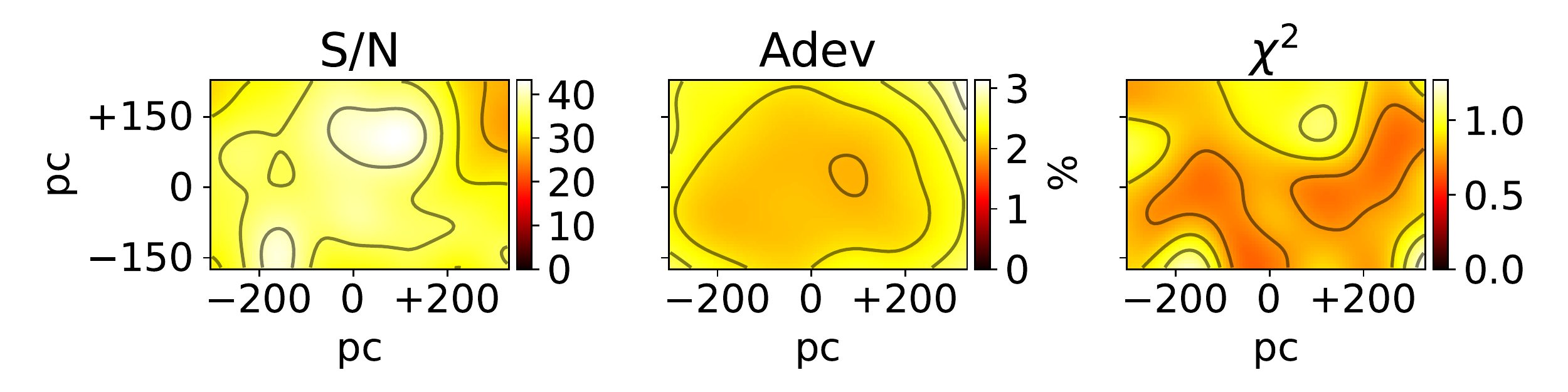}
    \caption{Maps displaying the statistical parameters derived from the stellar population synthesis. From left to right: the signal-to-noise ratio (measured between 5670-5730~\AA), Adev and $\chi^2$. Together they attest to the robustness of the modelled population.}
    \label{fig:stats_syn}
    \end{figure*}

    We present an example of the fits for an individual, nuclear spaxel in Fig.~{\ref{fig:example}}. A good matching between the observed (black) and modelled (red) spectra can be seen. The quality of the fits over the full FoV is ensured by the signal-to-noise (S/N) ratio map and can be certified in the $\chi^2$ and Adev\footnote{Adev is the Allan deviation and serves as a quality indicator of the fit. It corresponds to the percentage mean $|O_\lambda-M_\lambda|/O_\lambda$ deviation over all fitted pixels.} maps (Fig.~{\ref{fig:stats_syn}}).

\begin{figure*}
    \centering
    \includegraphics[trim={0 0 7cm 0}, clip, width=0.95\textwidth, height=0.95\textheight, keepaspectratio]{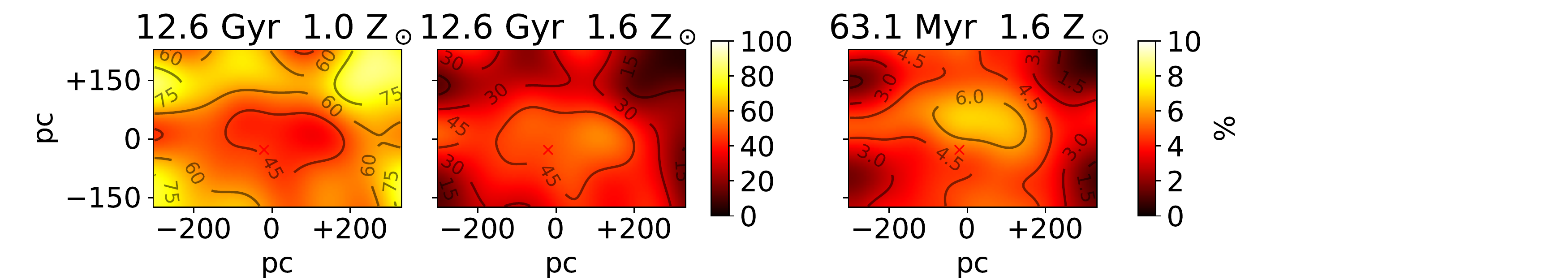}
    \caption{Maps displaying the light contribution from the three most significant components for the spectral synthesis in our FoV. The synthesis finds two old metal-rich (12.6~{Gyr}; 1.0 and 1.6~{Z$_\odot$}) components and a smaller contribution of a young also metal-rich (63.1~Myr; 1.6~Z$_\odot$) component. For display purposes, the scale of the younger population is different from that of the other two components.}
    \label{fig:components}
    \end{figure*}

    The resulting stellar population derived over the full FoV by the fitting procedure resulted in a contribution of mainly 3 components: two old metal-rich (12.6~{Gyr}; 1.0 and 1.6~{Z$_\odot$}) and a smaller contribution of a young also metal-rich (63.1~Myr; 1.6~Z$_\odot$). The spatial distribution of each component is shown in Fig.~{\ref{fig:components}}. It is clear that the central region is dominated by an old stellar population ($\sim12$~{Gyr}), as is to be expected for massive early-type galaxies. However, this contribution is divided into two different components with different metallicities, having 1.0 and 1.6~{Z$_\odot$} each. They also exhibit a distinct spatial distribution, with more metal-rich stars dominating the central region of our FoV. The same distribution is seen for the mass fractions derived for each component. The 1.0~{Z$_\odot$} SSP has a mean relative mass fraction of $\sim60$~{\%} ranging from $\sim41$~{\%} to $\sim84$~{\%}. On the other hand, the 1.6~{Z$_\odot$} SSP has a mean relative mass fraction of $\sim40$~{\%} ranging from $\sim14$~{\%} to $\sim58$~{\%}.
    
    \citet[][]{RickesEtAl2008} have also conducted stellar population synthesis studies in NGC\,6868 in a larger scale ( $R < 17$~{arcsec}) and also found an ubiquitous old population. However, they also report an intermediate-age stellar population of 5~{Gyr} which peaks in the centre of the galaxy, contradicting our findings. However, their data ranges from $5100 - 6800$~{\AA} therefore lacking the bluer end of the spectrum in order to constrain the presence of intermediate and young stars. Also, their SSP base consists of very few elements, that most likely do not portray all the different SFH a galaxy can have. Therefore, the discrepancies seen are most likely the result of better data employed here (larger wavelength coverage and S/N) and the improvement of the synthesis method.
    
    Despite the different FCs we tried in our base, we did not find any contribution from an AGN. As this is a galaxy classified as a LINER, the supermassive black hole is likely accreting at a really low rate, making its contribution to the continuum undetectable. 
    
\begin{figure}
    \centering
    \includegraphics[trim={0 0.5cm 0 0.5cm}, clip,width=\columnwidth, height=0.95\textheight, keepaspectratio]{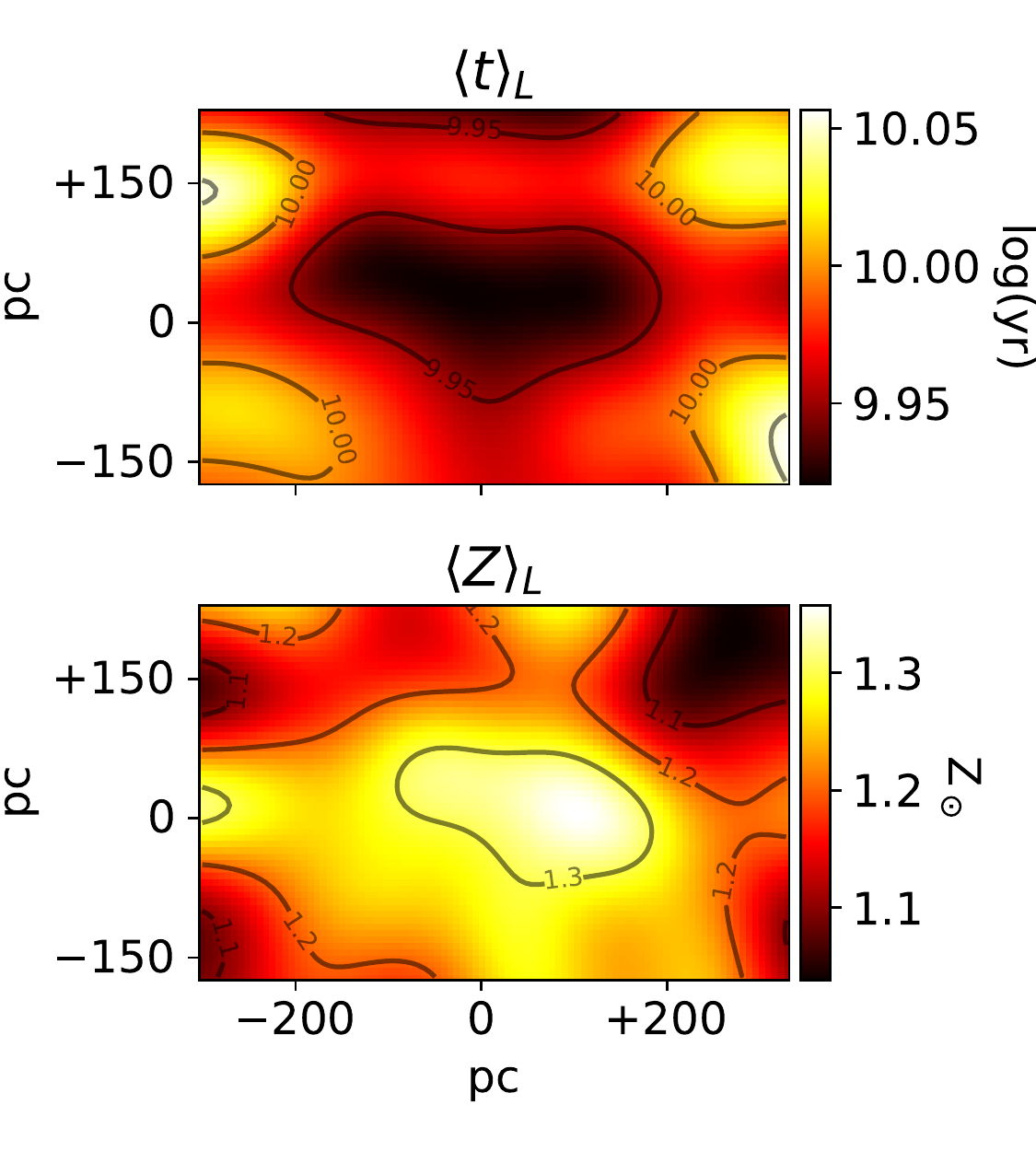}
    \caption{Mean age (left) and mean metallicity (right) maps derived from {\sc starlight} output. Both are luminosity-weighted quantities. The mean age seems to be affected by the younger component found in the synthesis and the mean metallicity map clearly shows the gradient metallicity in this galaxy. }
    \label{fig:mean_maps}
    \end{figure}

\begin{figure}
    \centering
    \includegraphics[trim={0 1cm 0 1cm}, clip,width=\columnwidth, height=0.95\textheight, keepaspectratio]{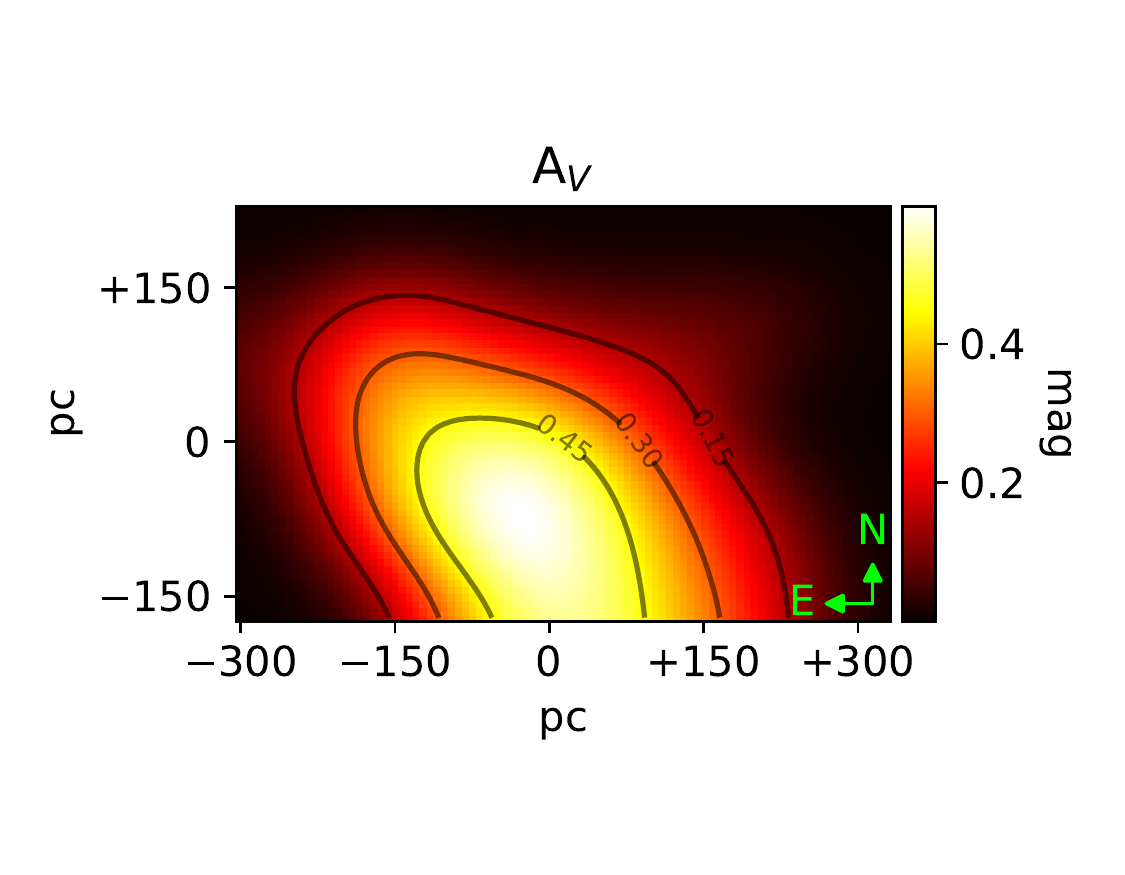}
    \caption{Map of the reddening in the V band extracted using {\sc starlight}. The morphology resembles a dust lane in the line of sight of the observer. This amount of dust has disturbed the continuum image, making it seem like a double core.}
    \label{fig:av_syn}
    \end{figure}

    In order to represent the galaxy's age and metallicity in a single map, we have derived the $\langle t \rangle_L$ and $\langle Z \rangle_L$ maps, which are shown in Fig.~{\ref{fig:mean_maps}}. In these maps, one can observe that the galaxy's mean age is slightly smaller and the metallicity higher in the nucleus. Finally in Fig.~{\ref{fig:av_syn}} we show the reddening map (A\textsubscript{V}). It reaches a peak of $\sim$0.65~{mag} in the centre of the image and its morphology resembles a dust lane embedded in the centre of the galaxy. This is in agreement with literature results \citep[e.g.][]{Veron-CettyVeron1988, HansenEtAl1991, BusonEtAl1993}.
    
\begin{figure}
    \centering
    \includegraphics[trim={0 0.5cm 0 0.5cm}, clip,width=\columnwidth, height=0.95\textheight, keepaspectratio]{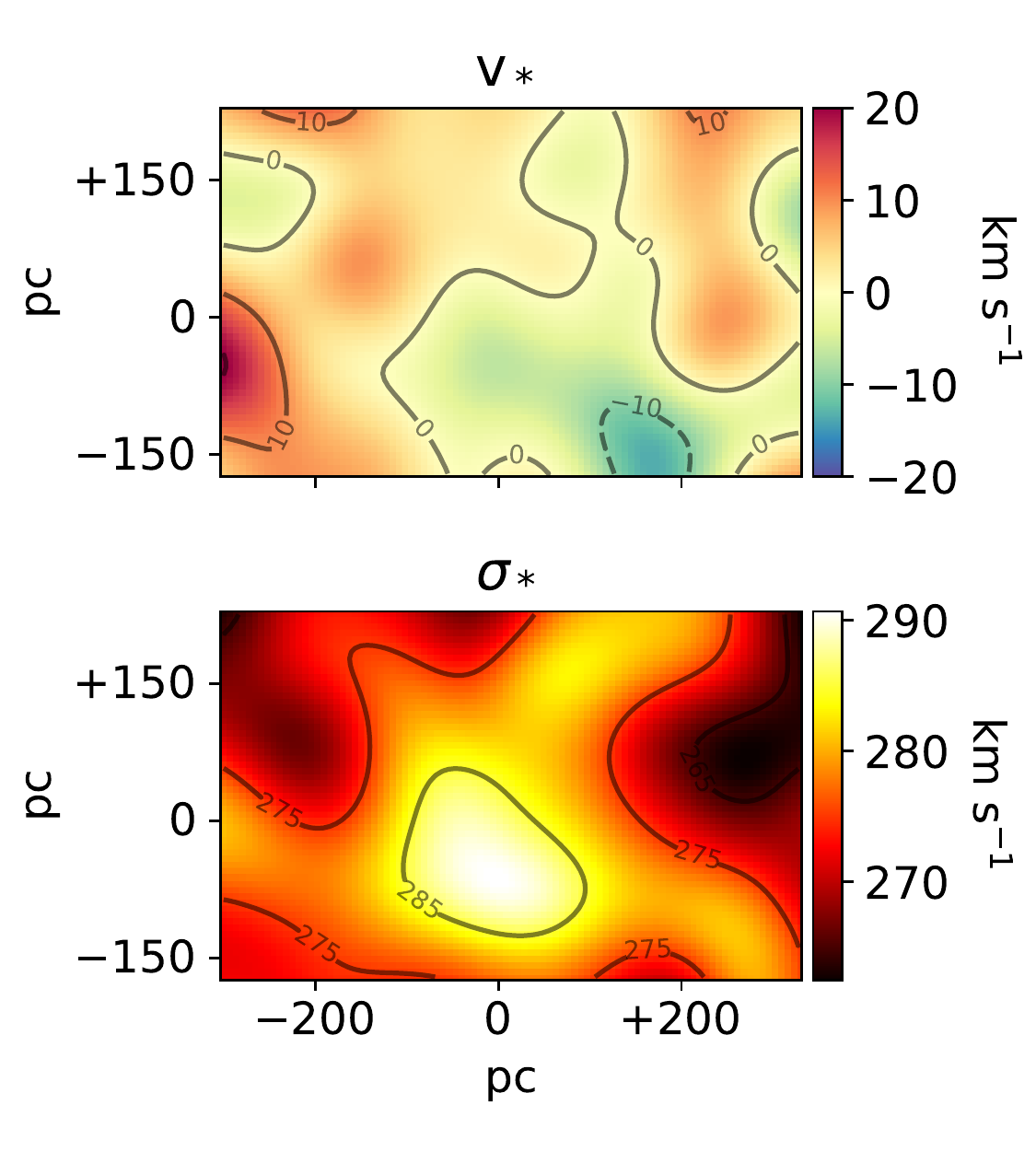}
    \caption{Kinematical maps regarding the LOSVD of the stellar component. The velocity map (left) does not seem to follow any particular geometry or distribution and the velocity dispersion (right) has a more defined profile with a clear peak at the centre of the distribution.}
    \label{fig:star_kin}
    \end{figure}

    {\sc starlight} also outputs the line of sight velocity and velocity dispersion. These maps are shown in Fig.~{\ref{fig:star_kin}}. It is evident that NGC\,6868 does not display a rotation profile or any ordered motion. Instead, it appears that the stars are in random motions, as can be seen in the $\sigma_*$ distribution, showing a clear peak $\sim 290$~{km s\textsuperscript{-1}} in the centre of the galaxy. In order to verify our results, we have also performed the fits using pPXF \citep{CappellariEmsellem2004, Cappellari2017} with the same SSP base and got the exact same results. Using the spectral resolution from our data and 5000~{\AA} as our reference wavelength, we come to the conclusion that we are not able to distinguish any velocities with less than $~100$~{km$\cdot$s\textsuperscript{-1}}. Therefore, we would need data with a better spectral resolution to properly characterize the kinematics of the central region of NGC\,6868.
    
    In past studies \citep[][]{CaonEtAl2000}, the stellar kinematics was measured using long-slit data and they found (1) a shallow rotation profile with a peak velocity of $\sim 45$~{km s\textsuperscript{-1}} at a radius of 42~{arcsec} ($\sim 5.8$~{kpc}); (2) a KDC which exhibits a counter rotation with respect to the outer regions and; (3) a drop in velocity dispersion at the centre of the galaxy, only seen in one of their PAs. However, in the inner part (<150 pc) of NGC6868  we are not detecting any sign of rotation nor drop in $\sigma_*$ reported. Moreover, comparing the derived velocity dispersion profiles from other PAs in their data, there is also no evidence of this behaviour. This is also the case for the KDC, where the velocities within it reach at most 10~{km s\textsuperscript{-1}}. The most probable explanation for these mixed findings is the high variation between different PAs as can be seen in Fig.~\ref{fig:star_kin}. Depending on the angle observed, a rotation (or counter-rotation) can or cannot be observed, due to the high-velocity dispersion in that region, making precise determinations of the velocity challenging. Moreover, an important caveat is that a KDC can only be detected when comparing the core and the outer parts of a galaxy. The FoV of our data does not allow such a comparison, therefore, in the present paper, we are unable to clearly say whether this object hosts a KDC or not.
    
    \subsection{Absorption line indices} \label{sec:res_idx}
    
\begin{figure*}
    \centering
    \includegraphics[width=1.0\textwidth, keepaspectratio]{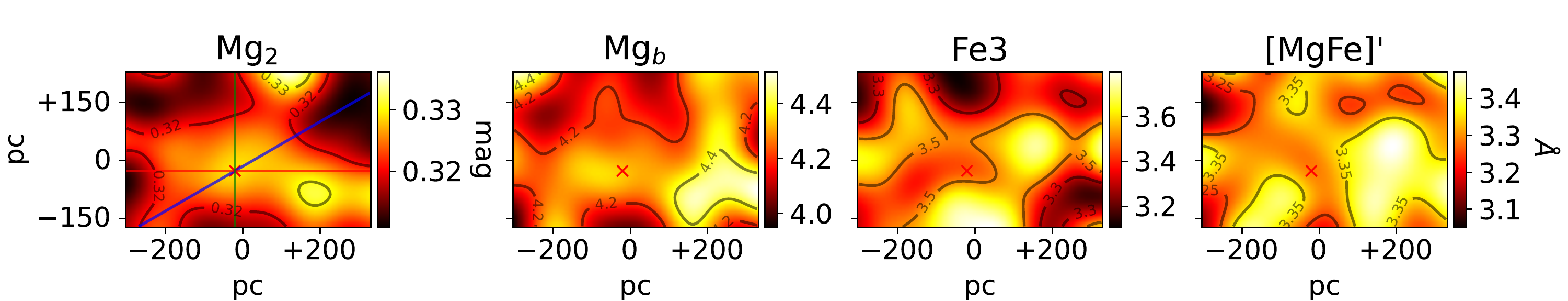}
    \caption{Results for the index measurement. From left to right: Mg\textsubscript{2}, Mg\textsubscript{b}, Fe3 and [MgFe]'. The indices show complex profiles in contrast to the results from the stellar population synthesis. In the Mg\textsubscript{2} map we over-plotted lines indicating the extractions done in our cube in order to compare with the literature results (Fig.~\ref{fig:other_studies}). The red "x" in the panels is the location of the peak Mg\textsubscript{2}.}
    \label{fig:idx}
    \end{figure*}

    The maps for the absorption line indices calculated in the Lick/IDS resolution are shown in Fig.~{\ref{fig:idx}}. All the values shown were corrected by the intrinsic velocity dispersion found with {\sc starlight}, except for the Mg\textsubscript{2} index because it is almost insensitive to Doppler broadening. In our case, this correction would be smaller than the intrinsic error of the method \citep[< 0.003, e.g.][]{Kuntschner2000}, so we chose not to apply any correction. This is not the case for the other indices. The correction factors for each spaxel follow the spatial distribution from the velocity dispersion (Fig.\ref{fig:star_kin}). The derived correction factors for the Mg\textsubscript{b} index are in the range 1.09-1.14; Fe4383, 1.12-1.17; Fe5270, 1.23-1.29; Fe5335, 1.40-1.54. The maximum error found for each index was Mg\textsubscript{2}: 0.0044~{mag}, Mg\textsubscript{b}: 0.15~{\AA}, Fe4383: 0.36~{\AA}, Fe5270: 0.14~{\AA}, Fe5335: 0.18~{\AA}, Fe3: 0.14~{\AA} e [MgFe]': 0.17~{\AA}. These errors leave all the following reported gradients unaffected.
    
\begin{figure}
    \centering
    \includegraphics[width=1.0\columnwidth, height=0.95\textheight, keepaspectratio]{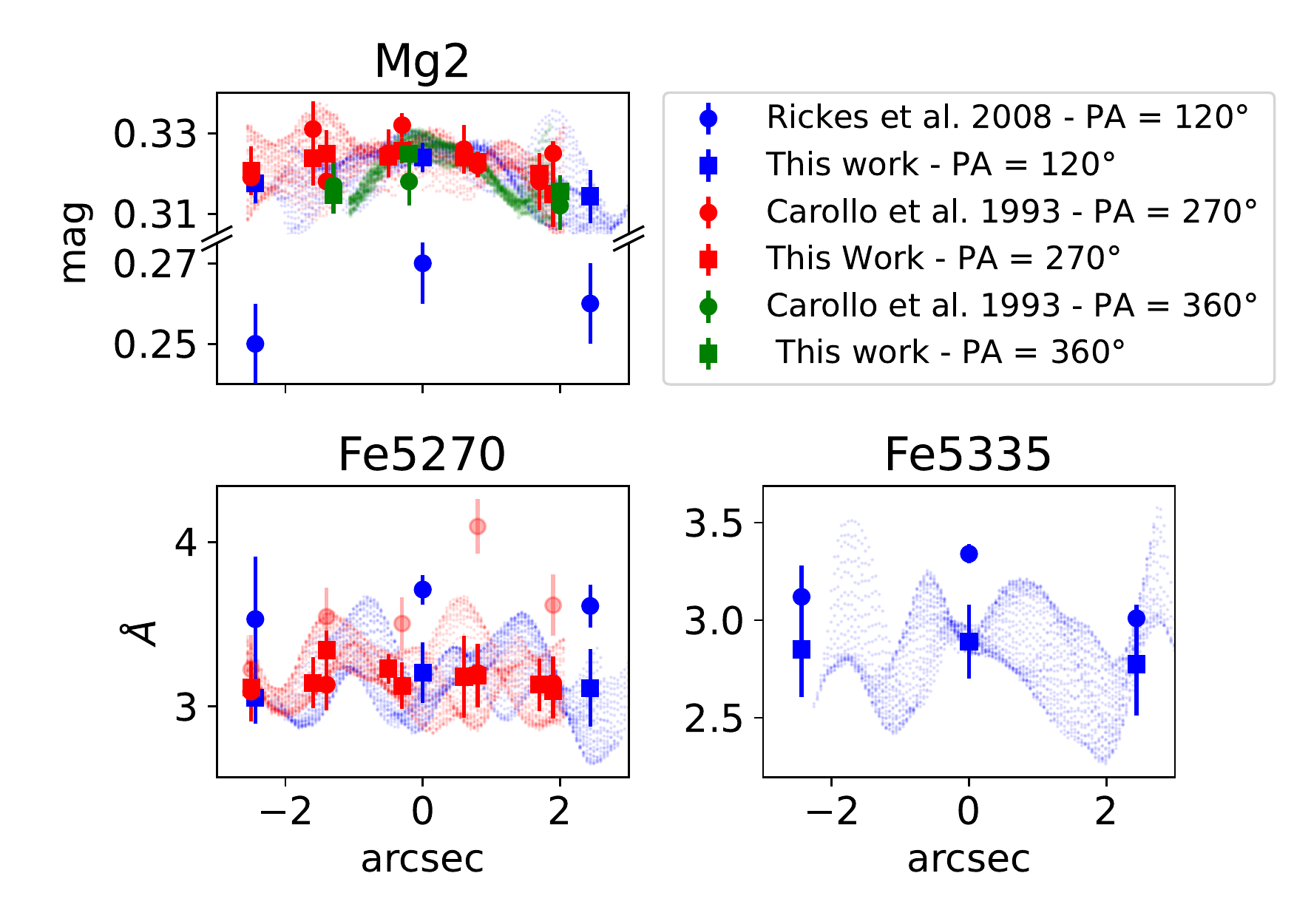}
    \caption{Comparison of our Mg\textsubscript{2}, Fe5270 and Fe5335 results to past studies using long-slit spectroscopy for different PAs \citep[][]{RickesEtAl2008, CarolloEtAl1993}. Each PA is represented by a colour, literature results are shown as circles and our measurements are shown as squares. The smaller scatter points correspond to the spaxels within the artificial slit observed. As can be seen, even inside the artificial slits there is significant variation.}
    \label{fig:other_studies}
    \end{figure}

    In order to assess the confidence of our results, we made extractions at PAs from the literature \citep[][]{CarolloEtAl1993, RickesEtAl2008} matching the spatial extent of our data. This comparison can be seen in Fig.~\ref{fig:other_studies} and the error bars displayed for the values derived in this work are the standard deviation measured within each bin which is greater than the systematic errors. Mg\textsubscript{2} presents an offset of at least -0.5~{mag} between \citet[][]{RickesEtAl2008} and our extractions, with both showing a negative gradient. The likely explanation for this offset is the different ways to measure the continuum applied in each work: they chose custom continuum bands and we followed \citet[][]{RiffelEtAl2019}. \citet[][]{CarolloEtAl1993}, on the other hand, matches all our data within the error bars for both of their PAs. Fe5270 and Fe5335 also present an offset between \citet[][]{RickesEtAl2008} and our measurements, again preserving gradients probably related to the continuum definitions, as previously said. \citet[][]{CarolloEtAl1993} measured Fe5270 (transparent points in Fig.~\ref{fig:other_studies}) and verified it was not in the Lick system. Therefore they derived a correction to be applied for their data in the form $\delta \text{Fe}5270 = 0.87(\pm0.07)\cdot \text{Fe}5270 - 2.67(\pm 0.02)$, resulting in $\delta \text{Fe}5270 \sim 0.5$~{\AA}. The corrected points are the full circles in Fig.~\ref{fig:other_studies} which, once again, match all our data within the error bars. \citet[][]{CarolloEtAl1993} did not measure Fe5335.
    
\begin{figure}
    \centering
    \includegraphics[width=\columnwidth, height=0.95\textheight, keepaspectratio]{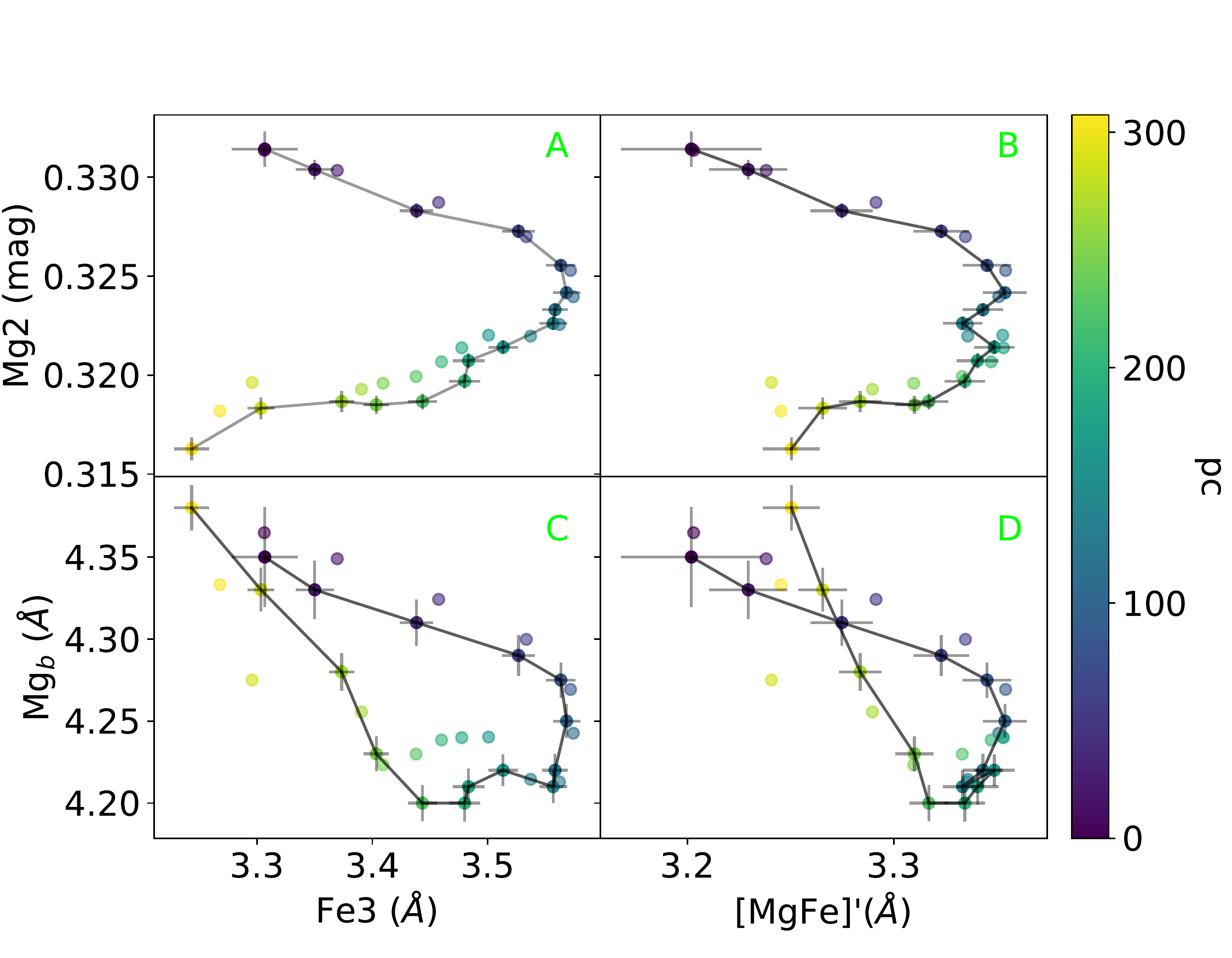}
    \caption{Correlations between Mg\textsubscript{2},  Mg\textsubscript{b}, Fe3 and [MgFe]' indices. These are the median values within $\delta r = 0.05$~{arcsec} apertures that have been colour-coded for the distance from the centre. For $R<100$~{pc} and $R>220$~{pc} the Mg\textsubscript{2} and Mg\textsubscript{b} anti-correlate , with Fe3 and [MgFe]'. The intermediate region displays a correlation. The transparent points indicate the mean of each index in each radial bin. This hints at different possible enrichment processes in each region. }
    \label{fig:profiles}
    \end{figure}

    In contrast to the parameters derived in the stellar population synthesis (Fig.~\ref{fig:mean_maps}), the spatial variation of the measured indices is far more intricate. Therefore to gauge the mean behaviour of each index, in Fig.~\ref{fig:profiles} we plot the median of each index in circular apertures of $\delta r = 0.05$~{arcsec} (7~{pc}) using the peak in Mg\textsubscript{2} as the origin. Mg\textsubscript{2} has the clearest spatial behaviour when compared to all the other indices, monotonically decreasing across the majority of the FoV. Two peculiar regions deviate from this trend, however: one at NW ~(+120 pc,+175 pc) in the border of the FoV, and another one SW at (+170 pc, -140 pc). Unfortunately, we do not have any literature results in these areas, nevertheless, given that our method resulted in an excellent agreement with past studies, we are convinced of this detection. The Mg\textsubscript{b} profile closely resembles Mg\textsubscript{2} also showing the distinct regions previously cited. Fe3 is particularly interesting because it does not follow the metallicity map as previously shown (Fig.~\ref{fig:mean_maps}), already a hint of \textalpha-enhancement processes at play. The most curious result is that in the same SW region previously mentioned, there is a depletion of this index. The [MgFe]' map closely resembles the $\langle Z_L\rangle$ map indicating that the centre of this galaxy really is metal-rich and exhibits a negative gradient.
    
    One curious behaviour that can easily be seen in Fig.~\ref{fig:profiles} is that for $R<100$~{pc} the Mg\textsubscript{2} and Mg\textsubscript{b} anti-correlate with Fe3 and [MgFe]' reaching their peaks at $R \sim 100$~{pc} away from the centre. Until $\sim220$~pc we actually observe a correlation between the two indices which turns again into an anti-correlation beyond $R \sim 220$~{pc}, however shallower when compared to the inner region.
    
    \citet[][]{RickesEtAl2008} found that Mg\textsubscript{2} and Fe5270 or Fe5335 establish a strong correlation which led them to conclude that the elements traced by both of these indices suffered the same enrichment processes. This can be the case for bigger scales, but in our FoV, we observe a distinct behaviour. 
    
    \subsection{Alpha-enhancement}
    
\begin{figure}
    \centering
    \includegraphics[trim={0.9cm 0 0 0}, clip,width=0.95\columnwidth, height=0.95\textheight, keepaspectratio]{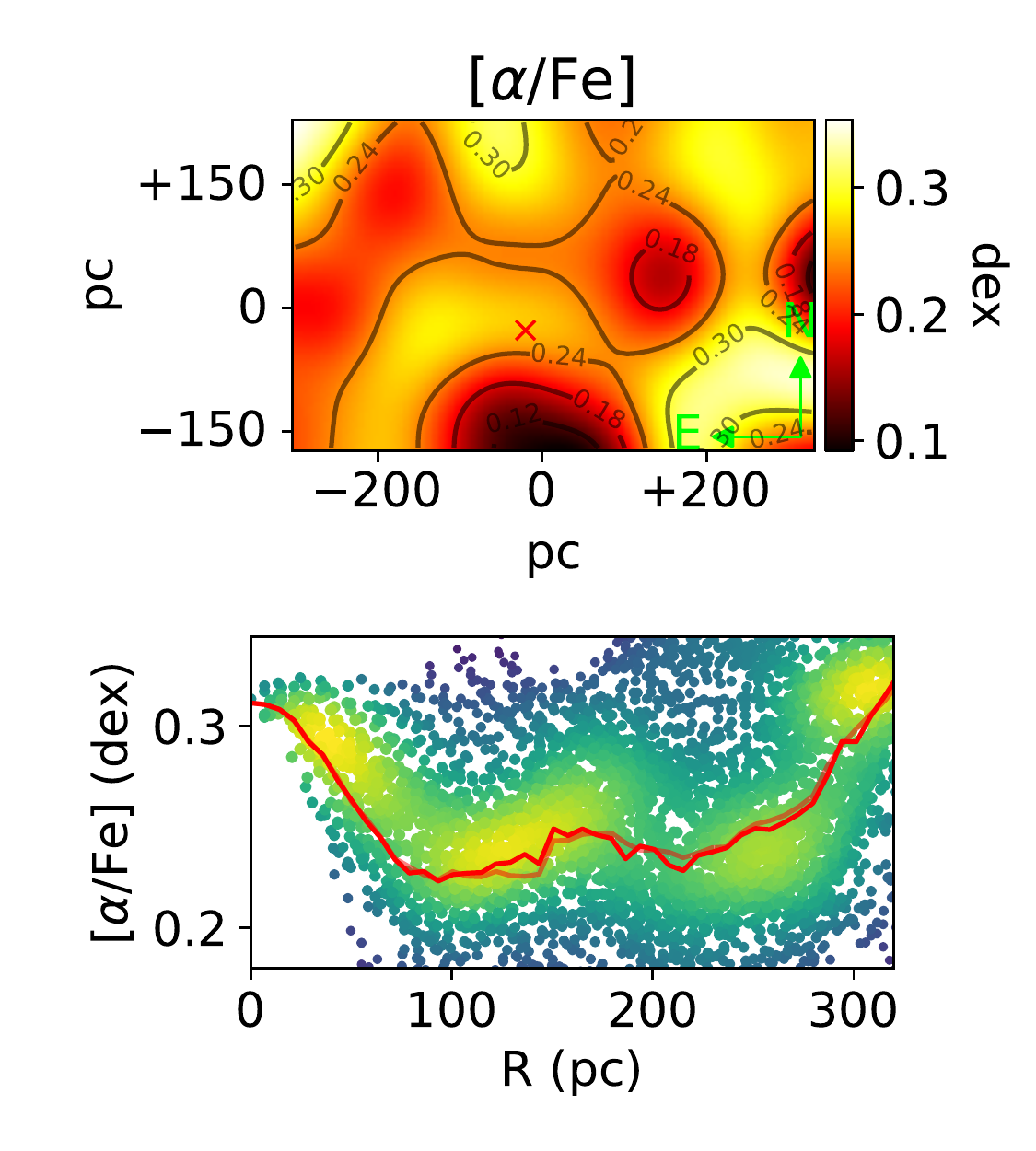}
    \caption{\textbf{Top panel:} Derived [\textalpha/Fe] for the whole FoV where it becomes clear that \textalpha-enhancement processes are ubiquitous. This parameter shows a really complex morphology, with a clear peak towards the centre, but also alpha-enhanced populations at larger radii. \textbf{Bottom panel:} Radial profile of [\textalpha/Fe] with the median (red) and mean (red, transparent) of the distribution over-plotted. It is clear that the profile exhibits a peak in the centre, followed by a region with smaller [\textalpha/Fe] followed again by an \textalpha-enhanced region. The red "x" in the top panel is the peak in Mg\textsubscript{2} following Fig.~{\ref{fig:profiles}} and is used as the reference to trace the profile in the bottom panel}
    \label{fig:aFe}
    \end{figure}

    The grids used to derive the [\textalpha/Fe] are seen in Fig.~\ref{fig:grids} as well as our measurements. As we only find old stellar populations, the most significant variation is given by the indices. Using these values, the [\textalpha/Fe] map is shown in Fig.~\ref{fig:aFe}. It is clear that the whole FoV presents \textalpha-enhanced stellar populations with values between $\sim+0.07$ and +0.24~{dex}. Also, we find a really structured profile with diverse morphology. Using once again the peak in Mg\textsubscript{2} as our reference, we plot the median profile of the [\textalpha/Fe] using circular apertures that can also be seen in Fig.~{\ref{fig:aFe}}. What becomes apparent is that the centre of NGC\,6868 shows a peak in \textalpha-enhancement followed by a region of shallower [\textalpha/Fe] which again is followed by another region with values as big as the ones in the centre, producing the "U" shape found in the median profile. 
    
    The only other paper where there is a measurement of the [\textalpha/Fe] from NGC\,6868 is \citet[][]{RickesEtAl2008}. According to them, the central parts of NGC\,6868 present lower [\textalpha/Fe], between -0.3 and 0.0~{dex} and an above-solar metallicity ([Z/Z$_\odot$]~$\sim+0.3$~{dex}. Moreover, the external parts present higher [\textalpha/Fe] values ($\sim+0.3$~{dex}) and lower metallicities ([Z/Z$_\odot$]~$\sim-0.33$~{dex}). In order to obtain these values, they use the Mg\textsubscript{2}, Fe5270 and Fe5335. However, as we already showed in \S~\ref{sec:res_idx}, these indices, despite agreeing with the gradients found in other studies, the values present a normalization problem probably due to the continuum bands used to compute the indices. If this shift in measurements were accounted for, the values of [\textalpha/Fe] would increase, reaching the \textalpha-enhanced region in their diagram, thus matching our observations. This would not affect the negative gradient in metallicity nor the positive gradient in [\textalpha/Fe] which are in agreement with observations of other early-type galaxies \citep[e.g.][]{KuntschnerEtAl2010}.

\section{Discussion} \label{sec:disscuss}

    \subsection{Stellar population synthesis}
    
\begin{figure}
    \centering
    \includegraphics[trim={0 1cm 0 1cm}, clip, width=0.95\columnwidth, height=0.95\textheight, keepaspectratio]{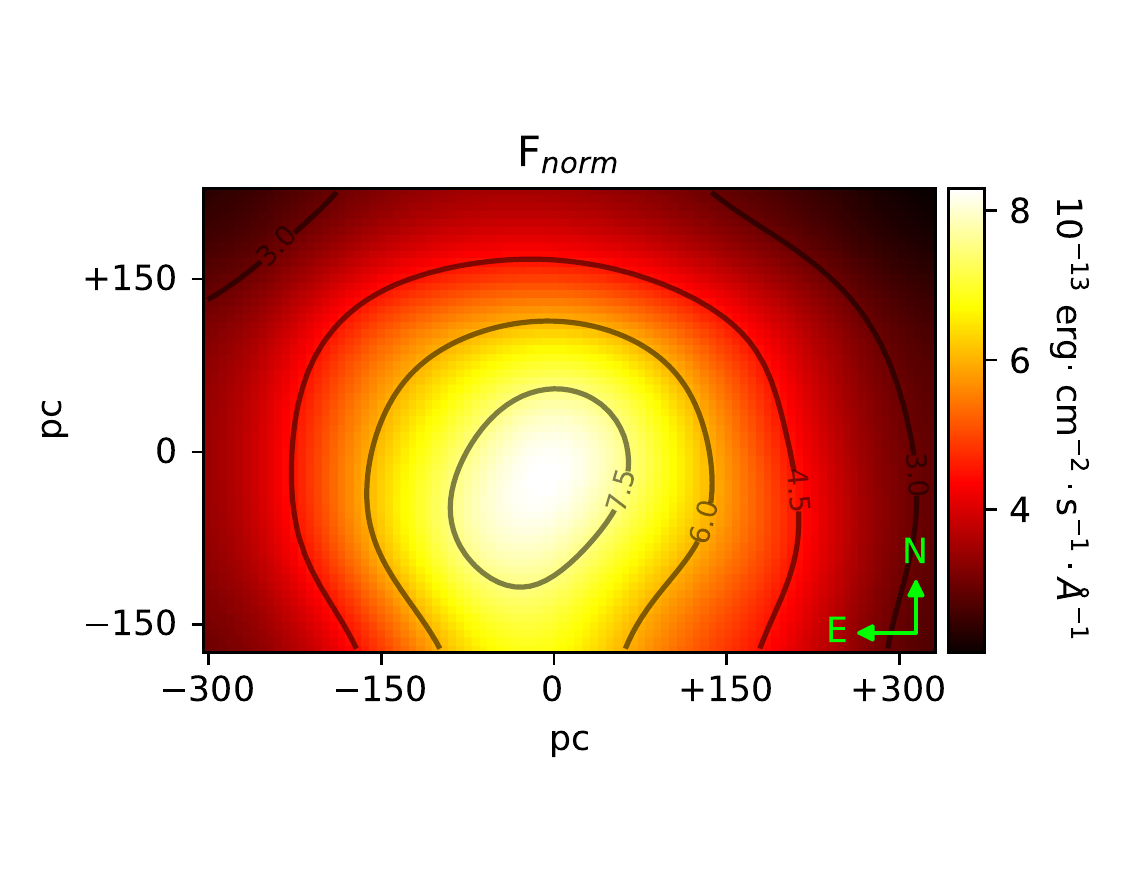}
    \caption{Continuum image extracted from our GMOS data of NGC\,6868. The actual photometric centre is uncovered after the correction by extinction.}
    \label{fig:fnorm_corr}
    \end{figure}

    NGC\,6868 is an early-type galaxy and so what is expected is that it had a fast phase of intense SF that suddenly stopped, forming the bulk of its stellar mass with a subsequent growth attributed to dry minor mergers. Therefore it is expected that the isophotes are not severely disturbed by the accreted galaxies. However, as we will present in the forthcoming paper (Benedetti et al. \textit{in preparation}), the photometric centre of NGC6868 has an offset with respect to the outer parts indicating a recent encounter with the NE dwarf companion galaxy. NGC\,6868 at first inspection appears to deviate from this hypothesis, exhibiting a significant distortion in its continuum image (Fig.~{\ref{fig:big_picture}}). When corrected by the stellar reddening found in the synthesis procedure, the actual morphology is revealed as an undisturbed spheroidal, as shown in Fig.~\ref{fig:fnorm_corr}.
    
    According to our analysis, the SFH of this galaxy is characterized by a short burst in the early Universe with no major episodes of star formation since then (Fig.~\ref{fig:mean_maps}). This means that NGC\,6868 probably did not experience any encounters with star-forming galaxies as they would leave signatures in the SFH from this galaxy. 

    The interpretation of the young metal-rich component (63.1~{Myr}; 1.6~{Z$_\odot$}) is tricky. For instance, it can be due to residual star formation which can be found in some elliptical galaxies. How common they are is still a matter of debate. \citet[][]{Salvador-RusinolEtAl2020} have detected young stars in massive red galaxies probably related to recycled material within the galaxy with mass fraction up to 0.5 per cent \citep[see also][]{deLorenzo-CaceresEtAl2020, Salvador-RusinolEtAl2021, Salvador-RusinolEtAl2022}. On the other hand, \citet[][]{SimonianMartini2017} interpreted early-type galaxies which typically are UV-weak as lacking younger components with HOLMES stars being responsible for this residual UV emission. Also, \citet[][]{BicaEtAl1996} using IUE spectra classified NGC\,6868 as a UV-weak source and found no contribution from young stellar populations.  In addition, \citet[][]{CidFernandesGonzalezDelgado2010} interpreted this young component as an artefact of the fitting process, and due to the lack of an old blue population, probably related to the horizontal branch, that current stellar population models do not account for.
    
\begin{figure}
    \centering
    \includegraphics[trim={0 1cm 0 1cm}, clip,width=0.95\columnwidth, keepaspectratio]{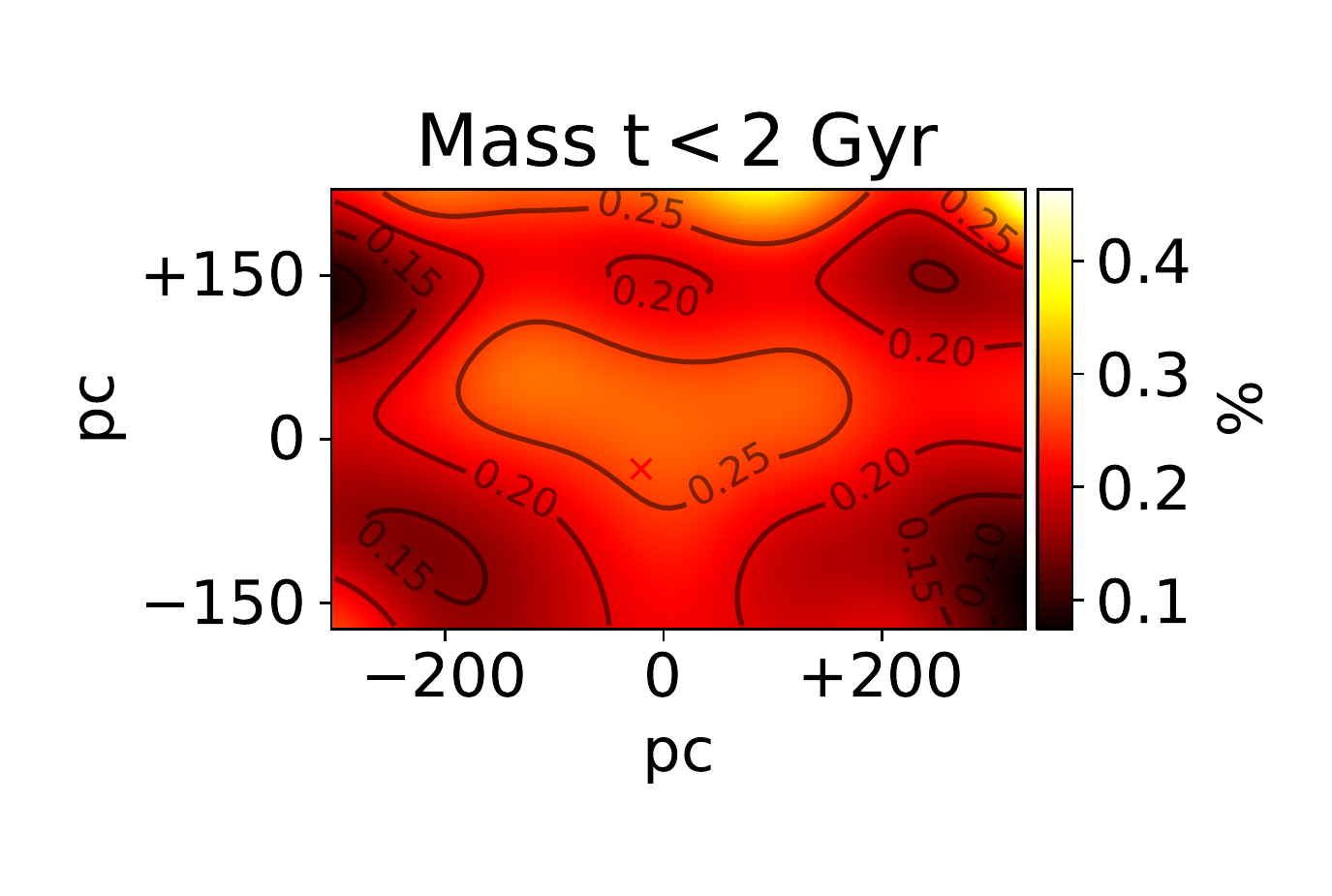}
    \caption{Map of the summed mass contribution found by {\sc starlight} from components younger than 2~{Gyr}. A clear increase towards the centre is seen, reaching 0.28~\% at the centre.}
    \label{fig:mass_2Gyr}
    \end{figure}

    In order to better understand the behaviour of this young component, we have used the M/L from the E-MILES models, and summing over the contributions from all SSPs with 2~{Gyr} or less, we get the map available in Fig.~\ref{fig:mass_2Gyr} with a median mass contribution in our FoV of $\sim0.2$~percent. There is a gradient for this component with higher values towards the nucleus of the galaxy \citep[the same behaviour as reported by][]{Salvador-RusinolEtAl2020}. In addition, this is a metal-rich (1.6\,$Z_{\odot}$) stellar population, thus most likely being formed from recycled material of former stellar generations.  Taking this all together, we interpret that the 63\,Myr contribution we found is due to a recent generation of stars, most likely formed from recycled gas from stellar evolution. However, it is worth mentioning that the present data do not allow us to completely rule out other mechanisms (e.g. HOLMES stars) that may also be enough to account for this fraction we found when fitting the data.

    It is worth mentioning that, as said in \S~\ref{sec:meth_syn}, we have tested for the presence of younger components in this object by using different SSP models that account for these younger stars in order to perform the stellar population synthesis. We found no evidence for the presence of younger components than 63~{Myr}.
    
    The mean metallicity map ($Z_L$, Fig.~\ref{fig:mean_maps}) is consistent with the scenario where massive early-type galaxies are able to retain the material expelled by SN, thus fixating the elements produced during stellar evolution into new stars, also creating a gradient where central regions tend to be more metal-rich when compared to outer regions. This is the case for the majority of early-type galaxies \citep[][]{KuntschnerEtAl2010}. They also show that for fast-rotators characterized by only old populations (>9~{Gyr}), a stellar disc is embedded in the stellar population of the galaxy characterized by higher metallicity and lower [\textalpha/Fe], including the central regions.
    
    The dust found in our data (Fig.~\ref{fig:av_syn}) is probably related to the cannibalism of a small gas-rich companion as proposed by \citet{HansenEtAl1991}. The molecular \citep[][]{RoseEtAl2019}, atomic \citep[][]{RoseEtAl2019} and ionized gas \citep[e.g.][]{BusonEtAl1993} detected are probably related with this event. For instance, as can be seen in figure 2 of \citet[][]{HansenEtAl1991}, there are dust filaments with spiral features that can be followed in a ring-like structure around the galaxy centre with a connection at NE which is aligned with the dust lane detected in the present study (Fig.~{\ref{fig:av_syn}}). The ring is also connected to a stellar-like object in NW that they proposed is a cannibalised galaxy. In fact, we found some recent, metal-rich, star-formation in this galaxy, as pointed out in \S~\ref{sec:res_synth} this residual star formation is most likely formed from material ejected from the previous generation of stars. In fact, the triggering/enhancing of AGN activity/luminosity has been related to an extra amount of gas that is added to the regular flow and is most likely originated from stellar evolution processes \citep{RiffelEtAl2022}. This gas (from the filaments + stellar evolution mass-loss) is most likely the reason behind the triggering of the AGN which is producing the LINER emission that is detected in the centre of the galaxy. This however is highly speculative and a thorough analysis of the ionized gas component will be carried out in an upcoming paper (Benedetti et al. {\it in preparation}).
    
    In the synthesis, despite including an FC component and testing with different exponents, the code does not find any significant contribution of an AGN to the continuum. This however does not mean there is not an AGN in the centre of NGC\,6868 as the emission can be obscured by dust, or the SMBH might not be accreting matter at a sufficiently high rate. As was already pointed out, this galaxy presents AGN evidence from past studies in radio \citep[][]{SleeEtAl1994, MauchEtAl2003}. We already know this is a LINER source, therefore the SMBH cannot be accreting at high rates. However, it is already known that when comparing AGN detection in the radio and optical the ratio of detection is not a 1:1 relation \citep[][]{ComerfordEtAl2020}.
    
    The kinematic revealed by {\sc starlight} matches what is expected for a cD galaxy, having $\sigma_*$ above 250~{km$\cdot$s$^{-1}$}. We do not find any clear sign of rotation or strong ordered motion. Actually, the v$_*$ map displays values that are below our uncertainty (see \S~\ref{sec:res_synth}, Fig.~\ref{fig:star_kin}). It is clear that our whole FoV presents values for the velocity in the line-of-sight which we are not able to confirm, however, for completeness we decided to show it in Fig.~\ref{fig:star_kin}. The velocity dispersion, on the other hand, does not face this problem as all values derived are $>250$~{km$\cdot$s\textsuperscript{-1}}. Despite past studies reporting a KDC in this object, we cannot confirm this result due to our small FoV. What remains clear is the fact that the central region of NGC\,6868 is really dispersion dominated and no kinematically distinct structure is found.
    
    \subsection{Absorption line indices and alpha-enhancement}\label{sec:disc_idx}
    
    Analysing the profiles derived for the indices (Fig.~\ref{fig:idx}), it is clear that they are far more structures when compared to the ones derived in the synthesis (Fig.~\ref{fig:mean_maps}). Moreover, the variation in each index is heavily dependent on the PA one decides to look at. This hints at a more chaotic assembly history when compared to what is derived in the synthesis that shows only contributions from old metal-rich populations. The [MgFe]' resembles the $\langle Z \rangle_L$ profile probably because we were able to partially isolate the sensitivity to the \textalpha-enhancement processes from the synthesis.
    
    This hypothesis is further endorsed by Fig.~\ref{fig:profiles}, where at least three distinct chemically enrichment regimes are clear, with two anti-correlations (R$\lesssim100$~{pc} and R$\gtrsim220$~{pc}) and a correlation (R$\gtrsim100$~{pc} and R$\lesssim220$~{pc}), clearer in the Mg\textsubscript{b}-[MgFe]' plot. What this tells us, again is that only a simple monolithic collapse cannot explain all our findings, otherwise we would expect the elements sensitive to the indices measured would have followed similar enrichment processes, thus producing matching gradients. This appears to be the case for larger scales \citep[][]{RickesEtAl2008}, however, this is not true for our FoV. A deviation from past studies that [MgFe]' reveals is a slight dip in the metallicity of the galaxy towards the very centre (R<100~pc, Fig.~\ref{fig:profiles}) that is also apparent in Fe3. As this index is insensitive to [\textalpha/Fe] what we observe is that the central region is depleted in metals with respect to the outer (100~{pc} < R <220~{pc}) region. This behaviour is unexpected as shown by past studies \citep[e.g.][]{KuntschnerEtAl2010}.
    
    The fact that we are able to detect such gradients using our observations is another reason why detailed studies on stellar populations on ETGs like the one presented here are necessary to further understand these objects. As shown here, despite these gradients naturally appearing in our data, in long-slit studies, as presented in \citet[][]{RickesEtAl2008} they are not able to detect any sign of this effect as their analysis is restricted to one PA.
    
    One way of distinguishing the formation scenario, as noted by \citet[][]{CarolloEtAl1993} is measuring the Mg\textsubscript{2} gradient (d~Mg\textsubscript{2}/d~log~r). Using only r>1 arcsec we find a gradient of d~Mg\textsubscript{2}/d~log~r$\approx$-0.024, which is a shallow gradient, incompatible with the monolithic collapse scenario according to the authors. 
    
    The [\textalpha/Fe] map (Fig.~\ref{fig:aFe}) also presents a complex profile. However, looking at the median curve we see a much clear behaviour. The central (R<100~{pc}) and outer (R>260~{pc}) regions appear to be significantly more \textalpha-enhanced than the intermediate region. These regions agree with our findings from the indices. What becomes clear is that the signatures in the stellar populations of the centre of NGC\,6868 cannot be described only by a burst of star formation in the early universe with a passive evolution since then.

    In order to test if the dilution by young stellar component was affecting our [\textalpha/Fe] estimates, we followed the same procedure as described in \S~\ref{sec:meth_idx} and estimated the [\textalpha/Fe] measuring the indices only in the synthetic spectrum. We found that the effects of the young population are only able to change at a maximum of 0.04~{dex} the [\textalpha/Fe]  values. Therefore, we conclude that the young component is not able to explain the spatial variation found in Fig.~\ref{fig:aFe}.
    
    \subsection{Possible formation scenario for NGC 6868} \label{sec:disc_form}
    
    From our findings, NGC\,6868 presents compelling evidence that its assembly history was not just comprised of a single burst of star formation without any significant evolution ever since. The [\textalpha/Fe] map is especially suited to understanding this galaxy. A close inspection reveals that the regions where it is smaller are also the more metallic regions. \citet[][]{KuntschnerEtAl2010} found stellar populations with these exact same features (high metallicity and low [\textalpha/Fe]) in discs of fast-rotator early-type galaxies. Some of these discs show signs of recent star formation, however, they also find these characteristics in galaxies depleted of gas and dominated only by old stellar populations (<9~Gyr).
    
    Therefore, our hypothesis is that in the past NGC\,6868 could have suffered an episode of merger with another galaxy with lower [\textalpha/Fe], such as the ones previously described. This would explain why we see such structured maps in [\textalpha/Fe], the slight increase in metallicity outside from the centre and the absence of a detectable subsequent star formation episode or a lack of gradient in the mean age across our FoV. The fact that we do not observe any clear kinematical signature is probably because we are looking at a really small region and cannot compare with outer regions, despite a KDC being previously reported \citep[][]{CaonEtAl2000}. However, we do find an imprinted chemical signature. We emphasize that, in this hypothesis, the accreted galaxy could not have a mass comparable with NGC\,6868, because, despite different maps showing structured profiles, Mg\textsubscript{2} shows a (shallow) negative gradient. Therefore, this enrichment process must have disturbed the stars from NGC\,6868 only to a certain extent. Simulations could be used to test our predictions, however, this analysis is beyond the scope of this paper.
    
    Finally, this might not be the only disturbance NGC\,6868 has suffered. We notice some regions that stand out, mainly in the SW in the [\textalpha/Fe] map which displays a $\sim+0.25$~dex and, as can be seen in the Fe3 map (Fig.~\ref{fig:idx}) it is depleted in [Fe/H]. This can be a recently captured small galaxy, however, this is unlikely due to its scale and the region it is encountered. We would need more data to properly characterize this region

\section{Concluding remarks}\label{sec:conclusion}
     
     In this paper, we analysed GMOS-IFU data of the inner region of the ETG NGC\,6868 mapping for the first time the physical and chemical properties of the stellar content of this source. This, together with an absorption-line indices analysis has allowed us to constrain the assembly history of this object. Our results can be summarized as follows:

    
    \begin{itemize}
        \item This galaxy is dominated by an old metal-rich (12.6 Gyr; 1.0 and 1.6 Z$_\odot$) stellar population and presents a negative gradient in metallicity. This is further endorsed by the [MgFe]'. 
        \item We found a recent ($\sim$ 63\,Myr) metal-rich ($1.6Z_{\odot}$) stellar population in the center of the galaxy. We suggest that this component is  most likely due to stars being formed
        from recycled material of former stellar generations.
        \item The apparent distortion in the continuum image is due to a dust lane embedded in the centre of the galaxy and reaches a peak in A\textsubscript{V}$\sim0.65$~{mag}. This structure is coincident with the one found in other studies.
        \item No evidence of an FC continuum is found, probably meaning the AGN in the centre of NGC\,6868 is accreting at really low rates.
        \item The kinematics in the centre of NGC\,6868 is characterized by high dispersion velocities and no apparent circular motion of the stars is seen.
        \item The indices Mg\textsubscript{2}, Mg\textsubscript{b}, Fe3 and [MgFe]’ all present structured profiles, with Mg\textsubscript{2} presenting the steepest negative gradient. However, it is too shallow to support a formation history due to a monolithic collapse. 
        \item Three distinct regions can be found when cross-correlating the indices: anti-correlations for R$\lesssim100$~{pc} and R$\gtrsim220$~{pc} and a correlation for 100~{pc}$\lesssim $R$\lesssim220$~{pc}. This reveals different enrichment histories in these regions.
        \item The [\textalpha/Fe] map also does not present a clear gradient. However, the median appears to also show three distinct regions: the central (R<100 pc) and outer (R>260 pc) regions appear to be significantly more \textalpha-enhanced than the intermediate region.
    \end{itemize}
    
    These findings suggest that NGC\,6868 was not formed on a single collapse and has passively evolved since then. Rather we propose that it has suffered a past merge with another galaxy. This can explain the findings in the \textalpha-enhancement and the different regions in the indices correlations together with the stellar population synthesis ones, such as the metallicity gradient and ubiquitous old ages. We do not find evidence of a distinct kinematic component either because this merger supposedly happened too long ago or we would need a larger FoV to asses if this region really is a KDC as other studies have previously reported.
    
    
\section*{Acknowledgements}

We thank the anonymous referee for the very useful comments and suggestions that helped to improve the manuscript. We also thank Alexandre Vazdekis for the insightful discussions. This work was supported by Brazilian funding agencies Conselho Nacional de Desenvolvimento Cient\'{i}fico e Tecnol\'ogico  (CNPq) and Coordena\c{c}\~ao de Aperfei\c{c}oamento de Pessoal de N\'{i}vel Superior (CAPES) and by the \textit{Programa de Pós-Graduação em Física} (PPGFis) at UFRGS. JPVB acknowledges financial support from CNPq and CAPES (Proj. 0001). RR acknowledges support from the Fundaci\'on Jes\'us Serra and the Instituto de Astrof{\'{i}}sica de Canarias under the Visiting Researcher Programme 2023-2025 agreed between both institutions. RR, also acknowledges support from the ACIISI, Consejer{\'{i}}a de Econom{\'{i}}a, Conocimiento y Empleo del Gobierno de Canarias and the European Regional Development Fund (ERDF) under grant with reference ProID2021010079, and the support through the RAVET project by the grant PID2019-107427GB-C32 from the Spanish Ministry of Science, Innovation and Universities MCIU. This work has also been supported through the IAC project TRACES, which is partially supported through the state budget and the regional budget of the Consejer{\'{i}}a de Econom{\'{i}}a, Industria, Comercio y Conocimiento of the Canary Islands Autonomous Community. RR also thanks to Conselho Nacional de Desenvolvimento Cient\'{i}fico e Tecnol\'ogico  ( CNPq, Proj. 311223/2020-6,  304927/2017-1 and  400352/2016-8), Funda\c{c}\~ao de amparo \`{a} Pesquisa do Rio Grande do Sul (FAPERGS, Proj. 16/2551-0000251-7 and 19/1750-2), Coordena\c{c}\~ao de Aperfei\c{c}oamento de Pessoal de N\'{i}vel Superior (CAPES, Proj. 0001). TVR thanks CNPq for support under grant 306790/2019-0. MT thanks the support of CNPq (process 312541/2021-0) and the programme L’Oréal UNESCO ABC \textit{Para Mulheres na Ciência}. RAR acknowledges the support from Conselho Nacional de Desenvolvimento Cient\'ifico e Tecnol\'ogico  and Funda\c c\~ao de Amparo \`a Pesquisa do Estado do Rio Grande do Sul. ARA acknowledges Conselho Nacional de Desenvolvimento Científico e Tecnológico (CNPq) for partial support to this work through grant 312036/2019-1. JAHJ acknowledges support from FAPESP, process number 2021/08920-8.

Based on observations obtained at the international Gemini Observatory and processed using the Gemini {\sc iraf} package, a programme of NSF’s NOIRLab, which is managed by the Association of Universities for Research in Astronomy (AURA) under a cooperative agreement with the National Science Foundation on behalf of the Gemini Observatory partnership: the National Science Foundation (United States), National Research Council (Canada), Agencia Nacional de Investigación y Desarrollo (Chile), Ministerio de Ciencia, Tecnología e Innovación (Argentina), Ministério da Ciência, Tecnologia, Inovações e Comunicações (Brazil), and Korea Astronomy and Space Science Institute (Republic of Korea).

\section*{Data Availability}

The data are publicly available on {\sc gemini} archive under the project GS-2013A-Q-52.



\bibliographystyle{mnras}
\bibliography{zotero} 








\bsp	
\label{lastpage}
\end{document}